\documentclass[usenatbib,fleqn]{mnras}

\usepackage{mathptmx}

\usepackage{amsmath}

\usepackage{graphicx}

\usepackage{hyperref}

\usepackage{xspace}

\usepackage{color}

\newcommand{\dotp}[2]{\mathbf{#1}\cdot\mathbf{#2}}

\newcommand{\ft}[3]{\int{#1}e^{-i\dotp{#2}{#3}}\;\mathrm{d}^3{#3}}

\newcommand{\epow}[1]{\mathrm{e}^{#1}}
\newcommand{\Si}{\mathrm{Si}}
\newcommand{\Ci}{\mathrm{Ci}}
\newcommand{\average}[1]{\langle#1\rangle}

\newcommand{\Mpc}{h^{-1}\,\mathrm{Mpc}}

\newcommand{\iMpc}{h\,\mathrm{Mpc}^{-1}}

\newcommand{\Om}{\Omega_\mathrm{m}}

\newcommand{\Ob}{\Omega_\mathrm{b}}
\newcommand{\Ow}{\Omega_w}

\newcommand{\om}{\omega_\mathrm{m}}
\newcommand{\ob}{\omega_\mathrm{b}}

\newcommand{\dc}{\delta_\mathrm{c}}

\newcommand{\Dv}{\Delta_\mathrm{v}}

\newcommand{\eg}{e.g.,\xspace}
\newcommand{\ie}{i.e.\xspace}

\newcommand{\nbody}{$N$-body\xspace}
\newcommand{\LCDM}{$\Lambda$CDM\xspace}
\newcommand{\halofit}{\textsc{halofit}\xspace}
\newcommand{\pkann}{\textsc{pkann}\xspace}
\newcommand{\meadfit}{\textsc{hmcode}\xspace}
\newcommand{\emu}{\textsc{cosmic emu}\xspace}
\newcommand{\planck}{\textit{Planck}\xspace}
\newcommand{\wmap}{\textit{WMAP}\xspace}
\newcommand{\euclid}{\textit{Euclid}\xspace}
\newcommand{\cfhtlens}{CFHTLenS\xspace}
\newcommand{\dmonly}{\textsc{dmonly}\xspace}
\newcommand{\agn}{\textsc{agn}\xspace}
\newcommand{\refe}{\textsc{ref}\xspace}
\newcommand{\dblim}{\textsc{dblim}\xspace}

\newcommand{\meadaddress}{\url{https://github.com/alexander-mead/hmcode}\xspace}

\newcommand{\new}[1]{{\color{black}{#1}}}
\newcommand{\newtwo}[1]{{\color{black}{#1}}}

\def\mnras{MNRAS}
\def\apj{ApJ}
\def\apjl{ApJ Let.}
\def\apjs{ApJS}
\def\prd{Phys. Rev. D}
\def\aap{A\&A}
\def\jcap{J. Cosmol. Astropart. Phys.}

\def\m@th{\mathsurround=0pt }
\def\eqalign#1{\null\,\vcenter{\openup1\jot\m@th\ialign{\strut\hfil$\displaystyle{##}$&$\displaystyle{{}##}$\hfil\crcr#1\crcr}}\,}
\def\gtsim{\mathrel{\lower0.6ex\hbox{$\buildrel {\textstyle >}\over {\scriptstyle \sim}$}}}
\def\ltsim{\mathrel{\lower0.6ex\hbox{$\buildrel {\textstyle <}\over {\scriptstyle \sim}$}}}

\title[Accurate halo-model power spectra]{An accurate halo model for fitting non-linear cosmological power spectra and baryonic feedback models}
\author[A. J. Mead et al.]{A. J. Mead$^{1}$\thanks{E-mail:
am@roe.ac.uk}, J. A. Peacock$^{1}$, C. Heymans$^{1}$, S. Joudaki$^{2}$ and A. F. Heavens$^{3}$\\
$^{1}$Institute for Astronomy, University of Edinburgh, Royal Observatory, Blackford Hill, Edinburgh EH9 3HJ\\
$^{2}$Centre for Astrophysics \& Supercomputing, Swinburne University of Technology, PO Box 218, Hawthorn, VIC 3122\\
$^3$Imperial Centre for Inference and Cosmology (ICIC), Imperial College London, Blackett Laboratory, Prince Consort Road, London SW7 2AZ\\
}

\date{Accepted 2015 September 01.  Received 2015 August 27; in original form 2015 May 28.}
\pagerange{\pageref{firstpage}--\pageref{lastpage}} 
\pubyear{2015}

\begin{document}
\maketitle

\label{firstpage}

\begin{abstract}
We present an optimized variant of the halo model, designed to produce accurate matter power spectra well into the non-linear regime for a wide range of cosmological models. To do this, we introduce physically motivated free parameters into the halo-model formalism and fit these to data from high-resolution \nbody simulations. For a variety of \LCDM and $w$CDM models the halo-model power is accurate to $\simeq 5$ per cent for $k\leq 10\iMpc$ and $z\leq 2$. An advantage of our new halo model is that it can be adapted to account for the effects of baryonic feedback on the power spectrum. We demonstrate this by fitting the halo model to power spectra from the OWLS hydrodynamical simulation suite via parameters that govern halo internal structure. We are able to fit all feedback models investigated at the 5 per cent level using only two free parameters, and we place limits on the range of these halo parameters for feedback models investigated by the OWLS simulations. Accurate predictions to high-$k$ are vital for weak lensing surveys, and these halo parameters could be considered nuisance parameters to marginalize over in future analyses to mitigate uncertainty regarding the details of feedback. Finally, we investigate how lensing observables predicted by our model compare to those from simulations and from \halofit for a range of $k$-cuts and feedback models and quantify the angular scales at which these effects become important. Code to calculate power spectra from the model presented in this paper can be found at \meadaddress.
\end{abstract}


\section{Introduction}

In the standard theory of cosmological structure formation, all large-scale structure in the Universe forms via the gravitational collapse of small amplitude initial seed fluctuations.  This process results in a non-linear network of haloes, filaments and voids that is comprised of dark matter and baryons. One of the goals of modern cosmology is to probe these density fluctuations in the late Universe and to use them to constrain models of the cosmos. In the early Universe, or at very large scales today, the density fluctuations are small in magnitude and can be analysed using linear perturbation theory, which can be calculated precisely as a function of cosmological parameters -- including both baryonic and dark-matter components (\eg \citealt{Seljak1996}; \citealt{Lewis2000}; \citealt{Blas2011}). As structures evolve in the later Universe they grow and become non-linear. Various perturbative schemes have been developed in cosmology to analyse these fluctuations (see reviews by \citealt{Bernardeau2002}; \citealt{McQuinn2015}), which give insight into the onset of non-linear structure formation. However, the most successful cosmological probes to date focus on the regime of linear perturbations, for example baryon acoustic oscillations (BAO; \eg \citealt{Padmanabhan2012}) or the cosmic microwave background (\eg \citealt{PlanckXIII2015}). In future surveys, the Universe will be mapped in finer detail and in principle it will be possible to extract a great deal of information from non-linear perturbations.

Unfortunately, perturbative schemes fail as larger non-linearities develop, due to the inability of perturbation theory to model matter shell-crossing (see \citealt{McQuinn2015} for a 1D discussion). Bound structures in the Universe today consist of matter that has undergone many crossings and can represent large departures from the mean density. Currently, these extreme non-linearities can only be accurately modelled by running large cosmological simulations, which commonly assume collisionless matter, so-called dark-matter-only simulations. However, even accurate pure dark-matter simulations are computationally expensive and prohibit the wide space of possible cosmological parameters to be explored quickly. Furthermore, it can also be difficult to understand which physical processes are at work in yielding a given simulation output. Thus, an analytic model for the evolution of structure can be invaluable, both in terms of speed and of insight. In this paper, we use the halo model (\citealt{Peacock2000}; \citealt{Seljak2000}; \citealt{Cooray2002}). This has become established as an important tool for explanation, but it is known \emph{not} to provide the accuracy required for the interpretation of current data.

A key measure of scale-dependent inhomogeneity, which can be calculated via perturbation theory or from the halo model, but also measured in simulations, is the power spectrum of the density field. Based on analytical insights, calibrated with \nbody simulations, various approximate formulae for the non-linear power spectrum have been generated. The most widely used of these to date has been the \halofit method of \cite{Smith2003}, revised by \cite{Takahashi2012}, which uses ideas from the halo model. This fitting formula has been expanded by various authors; to include massive neutrinos (\citealt{Bird2012}) and $f(R)$ modified gravity models (\citealt{Zhao2014}).

A different approach that is used to make predictions for the non-linear spectrum is that of the `emulator' code based on the `Coyote Universe' suite of simulations (\citealt{Heitmann2009}; \citealt{Heitmann2010}; \citealt{Lawrence2010}; \citealt{Heitmann2014}): the \emu. Sets of high resolution simulations are run at key points in cosmological parameter space (so-called `nodes') so as to cover the space evenly. The emulator then interpolates between the measured power spectra as a function of cosmology, yielding predictions for any set of parameters within the space. \cite{Heitmann2014} show that their emulator produces the power spectrum to an accuracy of $1$ per cent for $k \le 1 \mathrm{Mpc}^{-1}$ and $5$ per cent for $k \le 10 \mathrm{Mpc}^{-1}$ and it covers a small, but interesting, range of cosmological parameter space for flat universes and dark energies with constant equation-of-state. However, \emu makes no predictions for $k>10\iMpc$ or $z>4$ and it would be useful to expand the range of cosmological parameters that it currently encompasses.

Even if one were in possession of a perfect model of non-linear gravitational clustering it is difficult to compare this theory directly with the total matter field in the Universe. Instead one typically views the matter field via its gravitational light-bending effect, specifically in how bundles of light from a distant galaxy are sheared as they pass density perturbations between the galaxy and a telescope. This gravitational shearing induces a correlation in apparent galaxy shapes on the sky, as light from galaxies that are close on the sky is distorted coherently. This `weak' gravitational lensing (see the review by \citealt{Kilbinger2015}) has been used to place constraints on cosmological models, for example COSMOS (\citealt{Schrabback2010}) or \cfhtlens (\citealt{Heymans2013}; \citealt{Kilbinger2013}) and there are many forthcoming surveys designed to measure this effect in finer detail. Weak lensing measures a projected version of the total matter distribution in the Universe, in which the same shear correlations at a given angle can be caused by a smaller scale density fluctuation close to the observer, or a larger-scale fluctuation further away. This mixing of scales means that theoretical predictions for 2D weak-lensing observables require predictions for the clustering of the full 3D matter distribution over a wide range of scales and redshifts -- although see the 3D lensing of \cite{Heavens2003} and \cite{Kitching2014} for ways to avoid this. Current lensing analyses that work at the level of shear correlations either employ \halofit (\eg \citealt{Schrabback2010}; \citealt{Heymans2013}; \citealt{Kilbinger2013}), emulator-based strategies (\citealt{Liu2015}; \citealt{Petri2015}) or the technique of simulation rescaling (\citealt{Angulo2010}; \citealt{Angulo2015}) to provide predictions for the matter spectrum for the required scales; these predictions are then converted to predictions for shear observables.


In this paper, we take a completely different approach to modelling the full non-linear spectrum; we present an optimised variant of the halo model that is able to predict the matter power spectrum accurately to wave-numbers of interest for current and future lensing surveys ($k\simeq 10\iMpc$). Our first goal is to provide accurate halo-model fits to dark-matter only simulations across a range of cosmological parameters; our approach is to identify parameters in the halo model that can be made to vary in a physically-motivated way and then to fit these to high-resolution simulated power spectra from the emulator presented in \cite{Heitmann2014}. This approach is distinct from that of \halofit (\citealt{Smith2003}; \citealt{Takahashi2012}), which is an empirical fitting formula motivated by the principles of the halo model, but one which does \emph{not} use the halo model directly. Although we focus on weak lensing observables we stress that our optimised version of the halo model is general, and useful for any cosmological analysis that currently uses \halofit. Our approach is also distinct from recent work aimed at improving the halo model by \cite{Mohammed2014a} and \cite{Seljak2015}. These authors do not use the full apparatus of the halo model in order to provide accurate matches to the power spectrum, instead opting to employ a combination of perturbation theory and a series expansion, aimed primarily at an accurate modelling of the quasi-linear regime. 


An additional source of uncertainty is the impact of baryonic feedback on the total matter distribution in the Universe. Most \new{treatments} of non-linear evolution ignore any interactions other than gravitational, except in the initial conditions. \new{Work with perturbation theory at large scales (\eg \citealt{Shoji2009}; \citealt{Somogyi2010}) has shown that including the distinct physics of dark matter and baryons offers small improvements ($\sim 0.5$ per cent in power) compared to the approximation of treating `matter' as being a single component. This has also been tested in simulations by \cite{Angulo2013a}, where it was shown that including distinct transfer functions for dark matter and baryons leads only to small differences in the eventual measured non-linear matter power spectrum -- below one per cent at late times around the BAO scale. In contrast to these small effects at large scales, early semi-analytical treatments (\citealt{White2004}; \citealt{Zhan2004})} and recent work using hydrodynamical simulations together with prescriptions for feedback (\eg \new{ \citealt{Jing2006}}; \citealt{Rudd2008}; \citealt{Schaye2010}; \new{ \citealt{Martizzi2014}}) have shown that the redistribution of matter caused by \new{non-linear} processes such as gas cooling, active-galaxy feedback and supernova explosions can have a large impact on the total mass distribution, but details regarding the magnitude of feedback are uncertain. We show that we are able to accurately capture the effects of a variety of feedback recipes using our optimised halo model by varying only two parameters that govern halo internal structure. We constrain these parameters and suggest limits for these that may form a prior. In forthcoming weak-lensing analyses these could either by constrained, to learn about feedback, or marginalized over to provide unbiased cosmological constraints.


This paper is structured as follows: In section~\ref{sec:halomodel} we first discuss our conventions for measures of inhomogeneity, then go on to discuss the halo model, including the specifics of the particular model that we employ. Next, in section~\ref{sec:modifications} we compare the original halo model to accurate \nbody simulations to display its shortcomings and then discuss our modifications to the model. In section~\ref{sec:results} we present our main result; the power spectra of the modified model which best fits the emulator of \cite{Heitmann2014}. We address baryonic feedback in section~\ref{sec:baryons} and demonstrate that our approach can be easily adapted to account for feedback in a variety of models. Finally in section~\ref{sec:lensing} we show how our predictions for the matter spectrum translate into lensing observables and show differences between our approach, that of \halofit, and simulations. Our work is then summarized in section~\ref{sec:conclusion}. Appendix \ref{app:response} is dedicated to showing the response of the halo-model power spectrum to variations in cosmological parameters while Appendix \ref{app:calculation} details the operation of our publicly available halo-model code.

\section{The halo model}
\label{sec:halomodel}

\subsection{Descriptions of inhomogeneity}

We will use the following notation: the matter density field, $\rho$, is given in terms of comoving position, $\mathbf{x}$, and time, $t$, by
\begin{equation}
\rho(\mathbf{x},t)=\bar{\rho}(t)[1+\delta(\mathbf{x},t)]\ ,
\end{equation}
where $\bar{\rho}(t)$ is the mean matter density and $\delta(\mathbf{x},t)$ is the density fluctuation about the mean. We will be interested in the Fourier Space overdensity $\delta_\mathbf{k}$, which is defined via the Fourier convention of \cite{b:Peebles} for a periodic volume $V$:
\begin{equation}
\delta_\mathbf{k} =\frac{1}{V}\ft{\delta(\mathbf{x})}{k}{x}\ ; \\
\label{eq:ft}
\end{equation}
\begin{equation}
\delta(\mathbf{x}) =\sum_\mathbf{k} \delta_\mathbf{k} e^{i\dotp{k}{x}}\ .
\label{eq:reverseft}
\end{equation}
The power spectrum of statistically isotropic density fluctuations depends only on $k=|\mathbf{k}|$ and is given by
\begin{equation}
P(k)=\langle|\delta_\mathbf{k}|^2\rangle\ ,
\label{eq:P(k)}
\end{equation}
where the average is taken over modes with the same modulus but different orientations. We find it more convenient to use the dimensionless quantity $\Delta^2$:
\begin{equation}
\Delta^2(k)=4\pi V\left(\frac{k}{2\pi}\right)^3 P(k)\ ,
\label{eq:Delta2_definition}
\end{equation}
which gives the fractional contribution to the variance per logarithmic interval in $k$. If the field is filtered on some comoving scale $R$, the variance is
\begin{equation}
\sigma^2(R)=\int_0^{\infty}\Delta^2(k)\, T^2(kR)\;\mathrm{d}\ln{k}\ ,
\label{eq:variance}
\end{equation}
where the window function is 
\begin{equation}
T(x)=\frac{3}{x^3}(\sin{x}-x\cos{x})\ ,
\label{eq:top_hat}
\end{equation} 
in the case of smoothing with a spherical top-hat.

\subsection{Halo-model power spectra}
\label{sec:original_hm}

The halo model is an entirely analytic model for the non-linear matter distribution in the Universe that takes inspiration from the results of \nbody simulations. The basic idea here goes back to \cite{Neyman1953} but  has been given a modern guise (\citealt{Seljak2000}; \citealt{Peacock2000}), as reviewed in \cite{Cooray2002}. The great power of the method is that it can encapsulate the clustering of galaxies by the choice of an appropriate halo-occupation number, specifying the number of galaxies as a function of halo mass. But the present application is simpler, since we are considering only the overall mass distribution.

We now summarise the main features of the method, following most closely the presentation in \cite{Peacock2000}: The density field is described as a superposition of spherically symmetric haloes, with mass function and internal density structure that are accurately known as functions of cosmology from simulations. In the simplest case of randomly distributed spherical haloes the power spectrum has the form of shot noise, moderated by the density profile of the haloes:
\begin{equation}
\Delta_\mathrm{1H}^2(k)=4\pi\left(\frac{k}{2\pi}\right)^3\frac{1}{\bar{\rho}^2}\int_0^\infty M^2 W^2(k, M) F(M)\;\mathrm{d}M\ .
\label{eq:halopower}
\end{equation}
Here the power spectrum is calculated as an integral over all halo masses, of mass $M$, where $F(M)$ is the halo mass function (comoving halo number density in $\mathrm{d}M$) and $W(k,M)$ is the normalized Fourier transform of the halo density profile:
\begin{equation}
W(k,M)=\frac{1}{M}\int_0^{r_\mathrm{v}}\frac{\sin(kr)}{kr}\ 4\pi r^2\rho(r,M)\;\mathrm{d}r\ , 
\end{equation}
where $r_\mathrm{v}$ is the halo virial radius. On large scales, haloes are not randomly distributed and displacements of haloes with respect to one another require us to consider a `two-halo' term to the power.  For the matter distribution, this is approximately the linear-theory power spectrum
\begin{equation}
\Delta_\mathrm{2H}^2(k)=\Delta_\mathrm{lin}^2(k)\ .
\label{eq:2haloterm}
\end{equation}
An expression for the full halo-model power spectrum is then given by a simple sum of the terms
\begin{equation}
\Delta^2(k)=\Delta_\mathrm{2H}^2+\Delta_\mathrm{1H}^2\ .
\label{eq:halomodel}
\end{equation}

\subsection{Ingredients of the halo model}

To implement this model, we need to know the halo mass function and density profiles in order to evaluate the one-halo integral in equation (\ref{eq:halopower}).

The halo density profile is commonly described via the \citeauthor*{Navarro1997} (\citeyear{Navarro1997}; NFW) profile:
\begin{equation}
\rho(r)=\frac{\rho_\mathrm{N}}{(r/r_\mathrm{s})(1+r/r_\mathrm{s})^2}\ ,
\label{eq:nfw}
\end{equation}
where $r_\mathrm{s}$ is a scale radius that roughly separates the core of the halo from the outer portion and $\rho_\mathrm{N}$ is a normalization; in order to have a finite mass this profile must be truncated at the virial radius $r_\mathrm{v}$, within which the mean overdensity of the halo is $\Dv$. More recent work \citep{Navarro2004} has shown that halo profiles can be better fitted with Einasto profiles, which differ from NFW near the halo centre. However, the halo-model power calculation (equation \ref{eq:halopower}) depends on a self-convolution of the profiles and this smears out details of the halo centre. Thus we prefer to use the simpler NFW fit.

Simulated haloes need to be identified in a particle distribution and this is usually determined via a user-set overdensity threshold. Typically, a value of $\Dv=200$ is taken, which is loosely based on predictions from the spherical collapse model in an $\Om=1$ universe, although some authors use a value of $\Dv$ that varies with cosmological parameters in accordance with spherical model predictions (\eg \citealt{Bryan1998}). Once the over density threshold has been set and the halo mass measured the virial radius is no longer an independent parameter, and in order to conserve mass 
\begin{equation}
r_\mathrm{v}=\left(\frac{3M}{4\pi\Dv\bar{\rho}}\right)^{1/3}\ .
\label{eq:virial_radius}
\end{equation}
Note that this means that in general the $r_\mathrm{v}$ given by equation~(\ref{eq:virial_radius}) will be different from the halo radius one may independently measure from halo particles in a simulation. The normalization, $\rho_\mathrm{N}$, is set by the requirement that the spherical integral of equation~(\ref{eq:nfw}) gives the halo mass. The only free parameter in a fit to simulations is then $r_\mathrm{s}$, or equivalently the halo concentration $c\equiv r_\mathrm{v}/r_\mathrm{s}$. An implication of this is that the value of $c$ measured for simulated haloes depends on the halo definition used -- particularly the $\Dv$ criterion, the algorithm used to identify haloes and the scheme used for breaking up spurious haloes or unbinding particles (\eg \citealt{Knebe2011}).

Since the genesis of the NFW profile, a large number of relations between the concentration and mass of haloes have been developed. The general trend is that haloes of higher mass are less concentrated than those of lower mass, attributed to the fact that larger haloes formed in the more recent past and that the central density of a halo retains a memory of the cosmological density at its formation time. The original $c(M)$ relation proposed by \cite{Navarro1997} was shown to produce an incorrect redshift evolution by \cite{Bullock2001}, who provided an updated relation based around the concept of halo formation time. Around the same time a similar model by \cite{Eke2001} was introduced, which was intended to predict the correct $c(M)$ relation in the case of models with the same background cosmological parameters but different linear spectra, for example warm dark matter models compared to a cold dark matter (CDM) model. Lately focus has shifted to produce extremely accurate concentration--mass relations for the standard \LCDM cosmological model (\eg \citealt{Neto2007}; \citealt{Gao2008}) but these relations do not allow for general variations in cosmology. More recently, \cite{Prada2012} and \cite{Klypin2014} have suggested $c(M)$ relations that are `universal', in that they do not depend on cosmology other than via the function $\sigma(M)$ (equation \ref{eq:variance}). These relations predict that models with identical linear spectra should share a $c(M)$ relation, at odds with results from the concentration emulator of \cite{Kwan2013}, which produces a different relation for models with identical linear spectra but different growth histories (\eg a \LCDM model compared to a $w$CDM model at $z=0$ with identical $\sigma_8$).

We choose to use the relations of \cite{Bullock2001} because it was derived by fitting to a wide variety of cosmologies and also because their haloes were defined with a cosmology-dependent overdensity criterion, and therefore naturally adapt to the changes that we plan to make to the halo model in section~\ref{sec:modifications}. The $c(M)$ formula relates the concentration of a halo, identified at redshift $z$, to a formation redshift, $z_\mathrm{f}$, via 
\begin{equation}
c(M,z)=A\frac{1+z_\mathrm{f}}{1+z}\ , 
\label{eq:bullock_cm}
\end{equation} 
where the parameter $A$ is deduced by fitting to simulated haloes. The formation redshift is calculated by finding the redshift at which a fraction ($f$, also derived from simulated haloes) of the eventual halo mass has collapsed into objects, using the \cite{Press1974} theory:
\begin{equation}
\frac{g(z_\mathrm{f})}{g(z)}\sigma(fM,z)=\dc\ ,
\label{eq:z_coll}
\end{equation} 
where $g(z)$ is the linear--theory growth function normalized such that $g(z=0)=1$, $\sigma^2$ is the variance of the linear density field filtered on the scale of a sphere containing a mass $M$ (equation \ref{eq:variance}; $M$ is the mass enclosed in a sphere with radius $R$ in the homogeneous universe), and $\dc$ is the linear-theory collapse threshold. The value of $\dc$ is calculated from the spherical collapse model: $\dc\simeq 1.686$ for $\Om=1$, with a \emph{very} weak dependence on cosmology (see \citealt*{Eke1996} for flat models with $\Lambda$ and \citealt{Lacey1993} for matter-dominated open models). In \cite{Bullock2001} the parameters $A=4$ and $f=0.01$ were found from fitting the $c(M)$ relation to halo profiles over a range of masses and cosmologies.

For very massive haloes, equation~(\ref{eq:z_coll}) can assign a formation redshift that is less than the redshift under consideration, suggesting that the halo formed in the future. In our calculations we remedy this by setting $c=A$ if $z_\mathrm{f}<z$, although it makes very little practical difference to our power spectrum calculations.

It was shown in \cite{Dolag2004} and \cite{Bartelmann2005} that the $c(M)$ relations proposed in \cite{Navarro1997}, \cite{Bullock2001} or \cite{Eke2001} failed to reproduce the exact variations in concentration seen in models with identical linear-theory power spectra but different models of dark-energy. Differences in concentration arise because haloes form at different times in these models, despite having matched linear theory at $z=0$, and the exact form of this hysteresis was not being captured by existing relations (although the general trend \emph{is} captured by \citealt{Bullock2001}). \cite{Dolag2004} proposed a simple correction scheme that augments the \LCDM concentration for a model by the ratio of asymptotic ($z\rightarrow\infty$) growth factors of the dark-energy cosmology to the standard \LCDM one:
\begin{equation}
c_\mathrm{DE}=c_{\Lambda}\frac{g_\mathrm{DE}(z\rightarrow\infty)}{g_\Lambda(z\rightarrow\infty)}\ ,
\end{equation}
and we implement this correction in our incarnation of the halo model. The effect of dark energy on halo concentrations can be seen at the level of the power spectrum in \cite{McDonald2006}, in fig. 10 of \cite{Heitmann2014} and also in our Fig.~\ref{fig:cosmology_variation_0.0}. It can be seen at the level of measured halo concentrations using the $c(M)$ emulator or \cite{Kwan2013}. Because halo concentration affects small-scale power (equation \ref{eq:halopower}), a corollary of this is that the full non-linear spectrum will be different at small scales in different dark-energy models, even if they share an identical linear spectrum. Any scheme in which the calculation of the non-linear power depends solely on the linear power will thus fail to capture this detail.

The mass function of haloes (the fraction of haloes in the mass range $M$ to $M+\mathrm{d}M$) has been measured from simulations (\eg \citealt{Sheth1999}; \citealt{Jenkins2001}) and has been shown to have a near-universal form, almost independent of cosmology, when expressed in terms of the variable
\begin{equation}
\nu\equiv\frac{\dc}{\sigma(M)}\ .
\label{eq:nudefinition}
\end{equation}
The mass function can be expressed as a universal function in $f(\nu)$, which is related to $F(M)$ that appears in equation~(\ref{eq:halopower}) via
\begin{equation}
\frac{M}{\bar{\rho}}F(M)\;\mathrm{d}\emph{M}=\emph{f}(\nu)\;\mathrm{d}\nu\ .
\label{eq:mfconversion}
\end{equation}
This universality was predicted in an approach pioneered by \cite{Press1974} whereby the mass function was calculated explicitly by considering what fraction of the density field, when smoothed on a given mass scale, is above the critical threshold for collapse ($\dc$) at any given time. The expression that they calculated for the mass function is the Gaussian
\begin{equation}
f(\nu)=\sqrt{\frac{2}{\pi}}\,\mathrm{e}^{-\nu^2/2}\ ,
\end{equation}
but this is not a good fit to the mass function as measured in simulations; therefore we use the improved formula of \cite{Sheth1999}, which was an empirical fit to simulations:
\begin{equation}
f(\nu)=A\left[1+\frac{1}{(a\nu^{2})^p}\right]\mathrm{e}^{-a\nu^2/2}\ ,
\label{eq:STmassfunction}
\end{equation}
where the parameters of the model are $a=0.707$ and $p=0.3$. $A$ is constrained by the property that the integral of $f(\nu)$ over all $\nu$ must equal one, therefore $A\simeq 0.2162$. In \cite{Sheth1999} $\dc$ was taken to be $1.686$, independent of the cosmology.

\new{Universality in the halo mass function is an approximation and the \cite{Sheth1999} mass function is only accurate at the $\simeq 20$ per cent level (\eg \citealt{Warren2006}; \citealt{Reed2007}; \citealt{Lukic2007}; \citealt{Tinker2008}; \citealt{Courtin2011}). However, we do not attempt to use an updated non-universal mass function prescription because we value the large parameter-space coverage of \cite{Sheth1999} over a more accurate mass function tuned to only a small region of cosmological parameter space.}

\section{Optimizing the halo model}
\label{sec:modifications}

\subsection{\emu}
\label{sec:cosmic_emu}

\begin{table}
\caption{\emu ranges for the six cosmological parameters that are allowed to vary. Note that $\omega_i=\Omega_i h^2$. $\Ob$ and $\Om$ are the cosmological densities of baryons and \emph{all} matter respectively, $n_\mathrm{s}$ is spectral index of primordial perturbations, $h$ is the dimensionless Hubble parameter, $w$ is the constant dark energy equation of state and $\sigma_8$ is the standard deviation of perturbations measured in the linear field at $z=0$ when smoothed with a top-hat filter of radius $8\Mpc$. Note that each model is kept flat, so that a change in $\om$ at fixed $h$ entails a change in $\Ow$ (the cosmological density in dark energy with equation of state $w$). The currently favoured \planck cosmology \citep{PlanckXIII2015} lies close to the \newtwo{`node 0'} values and the \emu parameter space encompasses at least $5\sigma$ deviations about this. Power spectra can be produced by the emulator between $z=0$ and $4$ and for scales from $k\simeq 0.002h$ to $10\iMpc$.}
\begin{center}
\begin{tabular}{c c c c c}
\hline
Parameter & Fiducial & Minimum & Maximum & \newtwo{Node 0}\\
\hline
$\ob$ & $0.0225$ & $0.0215$ & $0.0235$ & \newtwo{$0.0224$}\\
$\om$ & $0.1375$ & $0.120$ & $0.155$ & \newtwo{$0.1296$}\\
$n_\mathrm{s}$ & $0.95$ & $0.85$ & $1.05$ & \newtwo{$0.97$}\\
$h$ & $0.7$ & $0.55$ & $0.85$ & \newtwo{$0.72$}\\
$w$ & $-1$ & $-1.30$ & $-0.70$ & \newtwo{$-1$}\\
$\sigma_8$ & $0.755$ & $0.616$ & $0.9$ & \newtwo{$0.8$}\\
\hline
\end{tabular}
\end{center}
\label{tab:emu_parameters}
\end{table}

In this paper, we create an accurate halo model by fitting the halo model to data from high--resolution cosmological simulations. The simulation data we use comes from the \emu, originally described in \cite{Heitmann2010} and updated in \cite{Heitmann2014}. In these works, the authors ran a suite of high-resolution cosmological \nbody simulations and measured the power spectrum in each. The simulations encompass a range of cosmological parameters ($\ob$, $\om$, $n_\mathrm{s}$, $h$, $w$, $\sigma_8$) given in Table~\ref{tab:emu_parameters} and were constructed to fill the parameter space in an `orthogonal Latin hypercube' design to ensure that interpolating between models was as accurate as possible while still allowing only a small number (37) of simulations to be run. \emu is the code released by the collaboration and allows the power to be produced at any point in their parameter cube from $z=0$ to $4$. The full spectra produced by \emu are combinations of second--order Eulerian perturbation theory calculations at large scales and measured non-linear power at small scales. Extensive \new{simulation} resolution testing was conducted \new{(\citealt{Heitmann2010}; \citealt{Heitmann2014})}, and an accuracy of their \new{simulations} of $1$ per cent in $\Delta^2(k)$ to $k=1\iMpc$ and $5$ per cent to $10\iMpc$ is quoted. \new{Note that this is different from the accuracy of the interpolation scheme, which in the $h$-free version discussed in \cite{Heitmann2014} can be seen to be $\sim 3$ per cent accurate to $k=1\iMpc$ (degraded from the original $1$ per cent at $k=1\iMpc$ when $h$ is set free; see their figs 8 and 9) and $\sim 5$ per cent accurate to $10\iMpc$.}

\new{Should we trust the accuracy stated for \emu? Note that in answering this question it is important to specify whether one is comparing to the accuracy of the node simulations or of the emulator interpolation method. The nodes of \emu are compared to simulations in \cite{Takahashi2012} where it is shown that a subset of the simulations of \emu (\textsc{m001} to \textsc{m009} in \citealt{Lawrence2010}) seem to have systematically $\sim 2$ per cent less power at $k=1\iMpc$ when compared to the simulations presented in \cite{Takahashi2012}. However, the authors do not present their own resolution tests, and resolution issues may plausibly be the origin of this small discrepancy. In \cite{Smith2014} the accuracy of power spectrum predictions was investigated as a function of `non-physical' simulation parameters (such as force and mass resolution). The authors report that changing the \textsc{pmgrid} parameter in \textsc{gadget-2} (\citealt{Springel2005b}; the code used for the \emu simulations) can have the surprisingly large effect of $\sim 3$ per cent power differences at $k=1\iMpc$. This parameter was \emph{not} investigated in the resolution testing of \cite{Heitmann2010}. More recently \cite{Skillman2014} and \cite{Schneider2015} have shown disagreements around the $3$ to $5$ per cent level when comparing their own simulations to \emu for the \planck cosmology (\emph{not} one of the emulator nodes) for $k\leq10\iMpc$, but their claims are based on their own simulations of a single cosmological model and the error is within that quoted for the $h$-free \cite{Heitmann2014} extended emulator. Given this discussion we err on the side of trusting the stated simulation accuracy in the \emu papers with the caveat that the \textsc{pmgrid} claim of \cite{Smith2014} warrants further investigation and that additional comparisons to different suites of simulations would be beneficial.}

\new{We also note the existence of another power spectrum prediction code, \pkann (\citealt{Agarwal2012}; \citealt{Agarwal2014}), which uses neural networks to carry out interpolation between simulation nodes. We choose not to use this because the simulations it was trained on are not as high resolution as those of \emu (limited to $k\leq1\iMpc$). In addition, the \pkann simulations contain some hydrodynamics and we wanted to initially focus only on dark matter. However, \pkann does allow the user to vary neutrino mass, which may be useful in further work.}

Although having accurate power spectra is extremely useful, three obvious limitations of the \emu method are: How to extend the predictions to $k>10\iMpc$, or $z>4$? How to extend beyond the emulator cosmological parameter space? And how to account for baryonic feedback? The first question arises in weak-lensing studies where predictions for lensing observables technically require integrals of $\Delta^2$ over \emph{all} $k$ (see section~\ref{sec:lensing}) but the main challenge in extending to smaller scales is to include the complications of baryon feedback. However, modelling these scales is necessary if weak lensing is to achieve its claimed future precision. Additionally, in Monte Carlo Markov chain (MCMC) analyses, the chains will inevitably wish to explore outside the parameter range of \emu and it is unclear how to proceed in this case. In this paper, we fit a variant of the halo model to data from the `nodes' of \emu, which are the exact locations within the cosmological parameter space where the simulations were run. This has two advantages: The accuracy of the emulator is likely to be highest at the nodes, because there is no interpolation taking place. Secondly, in using the nodes we are taking advantage of the Latin hypercube deign of \emu. Our resulting halo model fits can be used to extend the simulations to higher $k$, or higher $z$, in a physical way because they are motivated by theoretical arguments. Additionally, we show in section~\ref{sec:baryons} that the halo model can be adapted to account for the influence of baryons on the matter power spectrum, by fitting parameters relating to halo internal structure to data from hydrodynamical simulations. We also suggest that a successful fitting recipe can be used directly to explore models outside the \emu parameter space. 

\begin{figure}
\begin{center}
\includegraphics[width=9cm,angle=270]{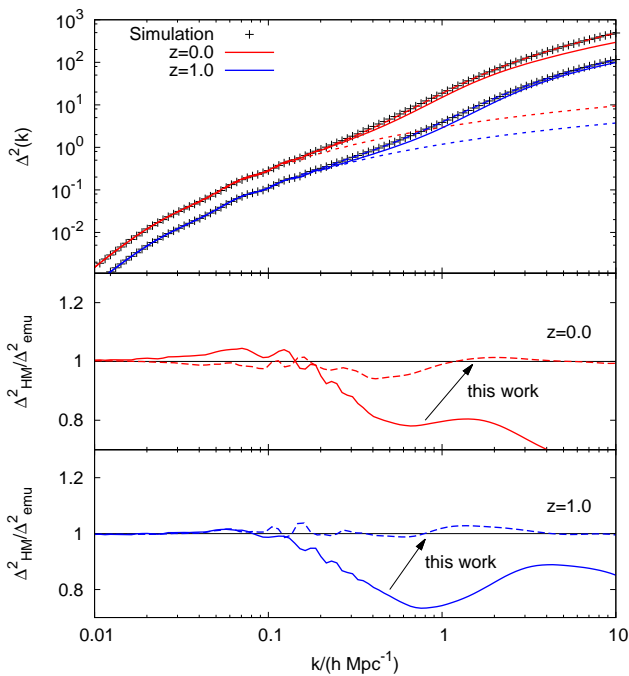}
\end{center}
\caption{A comparison of the original halo model, described in section~\ref{sec:original_hm}, to node 0 of \emu ($\Om=0.25$, $\Ob=0.043$, $n_\mathrm{s}=0.97$, $w=-1$, $\sigma_8=0.8$, $h=0.72$) at $z=0$ (upper; solid red), and $z=1$ (lower; solid blue). The spectrum from \emu is shown as black crosses. One can see that the halo-model power spectrum is qualitatively correct in shape, but under predicts the true power at the tens of per cent level at scales $k\gtsim0.2\iMpc$ at both redshifts. This is exactly the scale at which the one-halo term comes to dominate over the linear power (short-dashed lines in the top panel). The long-dashed lines show a preview of the final fits we are able to produce in this work, which agree with \emu at the $5$ per cent level across all scales. In the top panel, our model seems to track the simulation almost exactly and is hard to distinguish.}
\label{fig:original}
\end{figure}

In Fig.~\ref{fig:original}, we show a comparison of the power spectrum predicted by our original incarnation of the halo model (NFW haloes; \citealt{Bullock2001} concentrations; \citealt{Sheth1999} mass function; $\Dv=200$; $\dc=1.686$) to the power spectrum of \emu node 0, which is vanilla \LCDM near the centre of the parameter space, at $z=0$ and $1$. It is immediately obvious that the halo-model prediction is qualitatively reasonable in form, but deviates in detail from the simulations showing an underestimate of power of $\simeq 30$ per cent for $k>0.5\iMpc$. There are several possible reasons for the relatively poor performance of the halo model: Halo-finding algorithms tend only to assign half of the particles in a simulation into haloes (\citealt{Jenkins2001}; \citealt{More2011}) so the non-linear distribution of half of the mass in the simulation is treated by the halo model via an extrapolation of the formula for the mass function. \new{There are also clearly unvirialized objects in the quasi-linear regime that are not taken into account in our halo-model formalism, which also neglects halo substructure and asphericity as well as non-linear material that may lie outside the halo virial radius. In addition, a scatter in any halo property at fixed mass will change the predicted halo-model power spectrum. For example, \cite{Cooray2001} investigate a halo model with a scatter in $c(M)$, which typically boosts the power, while \cite{Giocoli2010} also include the power due to halo substructure via a substructure mass function. How the measured power spectrum is altered under various assumptions can also be seen in recent simulation work by \cite{vanDaalen2015} or \cite{Pace2015}.}

\new{Other problems are visible at large scales, where the halo model power can be seen to over-predict the simulations for $k<0.1\iMpc$. At these scales, the power is mildly non-linear and the two-halo term is in error, as well as the two- to one-halo transition. Attempts to accurately model quasi-linear scales using the halo model have been made by \cite{Valageas2011}, \cite{Mohammed2014a} and \cite{Seljak2015}, who use perturbation theory results as a two-halo term, and by \cite{Smith2007} who includes non-linear halo bias in the two-halo term. Accurate modelling of mildly non-linear power is an active field of research due to the importance of these scales for BAO measurements.}

\subsection{Fitting a general halo model}

\new{Rather than attempting to improve the halo model by adding missing ingredients (\eg \citealt{Smith2007}; \citealt{Giocoli2010}), thus making it more complicated, in this paper we take a more pragmatic approach:} it is possible that part of the inaccuracy of the power spectrum calculation stems partly from incorrect parameter choices. The model contains quantities such as $\Dv$, which are round numbers motivated by analytic arguments. We may therefore hope that improved results may be obtained by fitting the halo model to simulated power spectra using these quantities as physically-motivated free parameters. Our proposed changes represent a prescription for producing effective haloes whose power spectrum mimics the true one, even if these haloes differ from those measured directly in simulations. The hope is that we can trade off inaccuracies in \eg halo concentration against issues that are neglected in the standard halo model (asphericity, substructures, scatter in halo profiles), such that the two-point predictions are improved.

Nevertheless, we wish to retain the large amount of tested theoretical input that goes into the halo model. For example: changes in cosmological parameters alter the linear power spectrum, which in turn affects the mass function through the variance and the halo density profiles through the concentration and size relations. In addition, the linear growth rate will change, which also affects the concentration relations directly as well as the amplitude of the linear power spectrum. Since all of these ingredients have been tested against simulations, there are grounds for hoping that a small amount of parameter readjustment may allow the halo model to produce robust predictions for the non-linear power spectrum that are of useful accuracy for a wide range of cosmological parameters. The dotted lines in Fig.~\ref{fig:original} give a preview of how fruitful this approach is in providing an accurate model of the non-linear matter power spectrum.

\subsubsection{Adapting the two-halo term}

The two-halo term governs power on large scales and is given in its original form in equation (\ref{eq:2haloterm}). Linear theory slightly over-predicts the matter power spectrum around the quasi-linear scale and does a particularly poor job of modelling damping of the BAO peaks at $z=0$, which are damped by the quasi-linear effect of small-scale displacements. Modelling of the minutiae of the damping of the BAO peaks is beyond the scope of this work, but we treat the damping around these scales based on a model for the damping predicted from perturbation theory by \cite{Crocce2006b}, where
\begin{equation} 
\Delta^2_\mathrm{lin}(k)\to \epow{-k^2\sigma^2_\mathrm{v}}\Delta^2_\mathrm{lin}(k)\ ,
\label{eq:scoc}
\end{equation}
and $\sigma^2_\mathrm{v}$ is the 1D linear-theory displacement variance given by
\begin{equation}
\sigma^2_\mathrm{v}=\frac{1}{3}\int_0^\infty \frac{\Delta^2_\mathrm{lin}(k)}{k^3}\;\mathrm{d}\emph{k}\ .
\label{eq:sigma_v}
\end{equation}
The derivation of this expression assumes that the scales of interest are large compared to $\sigma_\mathrm{v}$, such that the damping factor cannot be trusted when $k\sigma_\mathrm{v}$ is large. We found that the best fit to numerical data at this scale required an expression equal to equation~(\ref{eq:scoc}) to quadratic order, but without the extreme high-$k$ truncation:
\begin{equation}
\Delta^{'2}_\mathrm{2H}(k)=\left[1-f\tanh^2{(k\sigma_\mathrm{v}/\sqrt{f})}\right]\Delta^2_\mathrm{lin}(k)\ ,
\label{eq:2halo_modified}
\end{equation}
where $f$ is a free parameter in our fit. In the $k\sigma_\mathrm{v} \gg 1$ limit, equation~(\ref{eq:2halo_modified}) reduces to $\Delta^2_\mathrm{2H}=(1-f)\Delta^2_\mathrm{lin}$. 

\subsubsection{Adapting the one-halo term}

\begin{figure*}
\begin{center}
\includegraphics[width=12.2cm,angle=270]{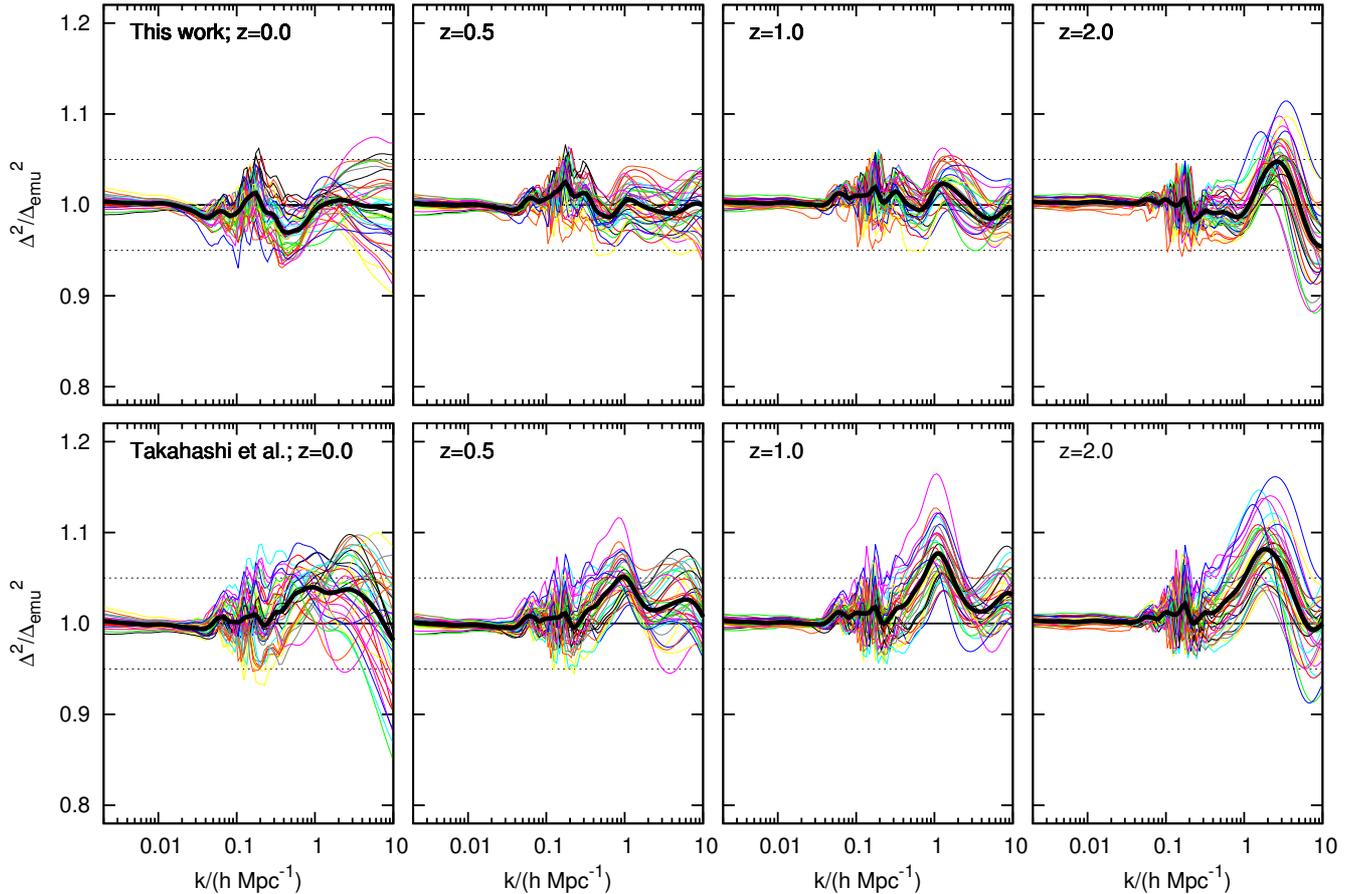}
\end{center}
\caption{Halo-model fits to all 37 nodes from \emu at $z=0$ (left) $0.5$, $1$ and $2$ (right) are shown in the top row, while the bottom row shows the predictions from the \citeauthor{Takahashi2012} (\citeyear{Takahashi2012}) revision of \halofit. The average fit is shown as the thick black line, whereas each individual thin coloured line shows a specific node.  The range of the coloured lines in each panel give an idea of how the accuracy of the model varies with cosmological parameters. The two models are of comparable accuracy for these cosmological models shown, although the halo-model approach performs slightly better. The halo-model fitting formula performs worst at $z=2$ where the high-$k$ power becomes systematically inaccurate and at $z=0$ where there is a spread in power at high $k$ values and a systematic under prediction in power around $k=0.3\iMpc$. However, it is mainly accurate at the 5 per cent level for all models and all scales shown. \emu itself is claimed to be accurate at the $1$ per cent level at $k=1\iMpc$ and $5$ per cent at $10\iMpc$, so our fit is mainly within this error around $k\simeq 10\iMpc$.}
\label{fig:best_fit}
\end{figure*}

We add freedom to the canonical form of the one-halo term in equation (\ref{eq:halopower}) in a number of ways: The first concerns the behaviour of the one-halo term at large scales, where the Universe tends to homogeneity faster than predicted by Poisson shot noise. At large scales, the one-halo term in equation~(\ref{eq:halopower}) decays as $\Delta^2\propto k^3$, whereas the linear power decays approximately $\propto k^4$, so it is inevitable that the one-halo term becomes greater than linear theory on very large scales, which is unphysical. This effect arises because haloes are treated as randomly placed in the standard halo-model formalism, when in fact they are clustered and distributed more smoothly than uniform random on very large scales. It has been suggested that a large-scale cut-off in the one-halo term can be physically explained as `halo exclusion' \citep{Smith2007} an effect that arises because, by definition, haloes cannot exist within each other. This is not captured by the standard halo-model power calculation because that calculation assumes that haloes are randomly placed, so that the probability of haloes overlapping is non-zero. Accounting for halo exclusion damps the halo-model power on large scales. Regardless of the exact details of exclusion, we modify the one-halo term so that it decays more rapidly than linear theory at large scales:
\begin{equation}
\Delta^{'2}_{1\mathrm{H}}=[1-\epow{-(k/k_*)^2}]\Delta^2_{1\mathrm{H}}\ ,
\label{eq:1halo_modified}
\end{equation}
where $\Delta^2_\mathrm{1H}$ is the same as in equation (\ref{eq:halopower}) and $k_*$ is a free parameter.

Within the one-halo term, parameters that we allow to vary are the virialized overdensity of a halo, $\Dv$, defined in equation~(\ref{eq:virial_radius}), and the linear collapse threshold, $\dc$, defined in equation~(\ref{eq:nudefinition}). Both of these parameters derive from the spherical model (\eg p. 488 of \citealt{b:peacock}) and rely on a somewhat arbitrary definition of the exact time of halo collapse. The variation of $\Dv$ can be predicted theoretically from the spherical model, and \cite{Bryan1998} provide a fitting formula\footnote{Equation~(\ref{eq:bn1}) differs slightly from that in \cite{Bryan1998} because we work with respect to the matter density, rather than critical density.} for a \LCDM cosmology
\begin{equation}
\Dv=\frac{18\pi^2+82[\Om(z)-1]-39[\Om(z)-1]^2}{\Om(z)}\ .
\label{eq:bn1}
\end{equation}
This suggests that $\Dv$ increases as the universe deviates from $\Om=1$.


\new{In standard theory, $\dc\simeq1.686$ but we allow this number to be a free parameter in our fit to power spectrum data. Note that changing $\dc$ changes the relationship between $\nu$ and the halo mass (equation \ref{eq:nudefinition}). This means that the `effective' mass function we invoke to improve $\Delta^2(k)$ predictions will \emph{not} necessarily accurately represent the mass function that might be measured in simulations.}

Fitted halo relations, such as the mass function and mass--concentration relation, depend upon how haloes are defined when identified in simulations. Therefore, the variations of $\Dv$ in our fitted halo model may not follow the simple theoretical variation in equation~(\ref{eq:bn1}) exactly, but we assume that the trend of increased $\Dv$ as the universe deviates from $\Om=1$ will serve as a useful initial guide when we explore parameter space. In addition, for flat models with a single component of dark energy it is expected that $\Dv$ would be a function of $\Om(z)$ only and this will be a useful principle in parameterizing fitting formulae. Increasing $\Dv$ has the effect of increasing the internal density of haloes and thus decreases the virial radius of a halo of a fixed mass, thus increasing small-scale power. Increasing $\dc$ means the linear density field has to reach higher values before collapse can occur (in the \citealt{Press1974} approach), the result of which is that the density field is dissected into more haloes of lower mass, which will reduce the amplitude of the shot-noise component of the one-halo term and thus reduce power.

One further free parameter is $\eta$, which we use to alter the halo window function via
\begin{equation}
W(k,M)\rightarrow W(\nu^\eta k,M)\ ,
\label{eq:eta_definition}
\end{equation}
changing the halo profile in a mass dependent way but leaving $\nu=1$ haloes unaltered \new{and the individual halo masses unchanged}. For $\eta>0$ higher mass ($\nu>1$) haloes are puffed out while lower mass haloes are contracted, both at constant virial radius: $\eta>0$ decreases the power whereas $\eta<0$ increases it. This extra ingredient was introduced to control the curvature of the power spectrum beyond $k\sim 1\iMpc$, where the filtering effect from the typical haloes has a major effect on the shape of the one-halo term. As we move to higher $k$ values, the properties of lower mass haloes become increasingly important. It is difficult for the one-halo term to track to the smallest scales, and correcting this requires an empirical perturbation of the halo profiles. Additionally, we allow ourselves to vary the amplitude of the concentration-mass relation: $A$ in equation~(\ref{eq:bullock_cm}).

\begin{table*}
\caption{Halo-model parameter descriptions and values before and after fitting}
\begin{center}
\begin{tabular}{c c c c c}
\hline
Parameter & Description & Original value & Fitted value & Equation in text \\
\hline
$\Dv$ & Virialized halo overdensity & 200 & $418\times\Om^{-0.352}(z)$ & \ref{eq:virial_radius} \\
$\dc$ & Linear collapse threshold & 1.686 & $1.59+0.0314\,\ln\sigma_8(z)$ & \ref{eq:nudefinition} \\
$\eta$ & Halo bloating parameter & 0 & $0.603-0.3\,\sigma_8(z)$ & \ref{eq:eta_definition} \\
$f$ & Linear spectrum transition damping factor & 0 & $0.188\times\sigma^{4.29}_8(z)$ & \ref{eq:2halo_modified} \\
$k_*$ & One-halo damping wavenumber & 0 & $0.584\times\sigma_\mathrm{v}^{-1}(z)$ & \ref{eq:1halo_modified} \\
$A$ & Minimum halo concentration & 4 &$3.13$ & \ref{eq:bullock_cm} \\
$\alpha$ & Quasi-linear one- to two-halo term softening & 1 & $2.93\times 1.77^{n_\mathrm{eff}}$ & \ref{eq:meadfit} \\
\hline
\end{tabular}
\label{tab:fit_params}
\end{center}
\end{table*}

\subsubsection{Full power}

A well known defect in the halo model is in the transition between the one- and two-halo terms, the so-called quasi-linear regime. In the standard halo model, the transition is modelled by a simple sum of the one- and two-halo terms (equation \ref{eq:halomodel}), but this is obviously deficient. At $z=0$, this transition scale is approximately $k=0.1\iMpc$ corresponding to physical scales of the order of tens of Mpc. On these scales, contributions to the density field will include, but are not limited to, large structure at the turn-around radius, sheets, filaments and voids. It would be rather surprising if the complexity of non-linear evolution on these scales could be accurately modelled by a simple sum of crude one- and two-halo terms. In testing, we noted that the halo model performed most poorly around these transition scales and we address this problem by modelling the transition via
\begin{equation}
\Delta^2(k)=[(\Delta_\mathrm{2H}^{'2})^\alpha+(\Delta_\mathrm{1H}^{'2})^\alpha]^{1/\alpha}\ ,
\label{eq:meadfit}
\end{equation}
where $\alpha$ is the final parameter that we adjust to match simulations. Values of $\alpha<1$ soften the transition between the two terms whereas $\alpha>1$ sharpen it. The power at these scales is quite smooth, so fitting the transition via $\alpha$ is sufficient. 

\section{Results}
\label{sec:results}

We fit the parameters introduced in the previous section to data from all 37 nodes of \emu at redshifts $z=0$, $0.5$, $1$, $1.5$ and $2$ with equal weight given to each redshift and node and $k$ weighted equally in logarithmic space from $0.01$ to $10\iMpc$. We use a least squares method to characterize goodness-of-fit and use an MCMC-like approach to fit all parameters simultaneously. Our best-fitting parameters are given in Table~\ref{tab:fit_params} where there are a total of $12$ parameters that are fitted to simulations, which can be compared with 34 for the \cite{Takahashi2012} version of \halofit. The cosmological dependences of each of our parameters was inferred by some experimentation. In Table~\ref{tab:fit_params}, we see that $\alpha$ depends on $n_\mathrm{eff}$, which is the effective spectral index of the linear power spectrum at the collapse scale, defined in \cite{Smith2003}:
\begin{equation}
3+n_\mathrm{eff}\equiv \left.-\frac{\mathrm{d}\ln\sigma^2(R)}{\mathrm{d}\ln R}\right\vert_{\sigma=1}\ .
\end{equation}
However, our $n_\mathrm{eff}$ is slightly different from that in \cite{Smith2003} because we define $\sigma(R)$ using a top-hat filter, rather than a Gaussian. 

The accuracy of this model is demonstrated in the upper row of Fig.~\ref{fig:best_fit}, which shows a ratio of the halo model to \emu at $z=0$, $0.5$, $1$ and $2$. One can see that our fitted halo-model predictions are mainly accurate to within $5$ per cent across all redshifts for the range of scales shown. We call this calibrated halo model \meadfit and refer to it thus throughout the remainder of this work. We also tested our model at $z=3$, a redshift to which it was not calibrated, and found that errors rarely exceed $10$ per cent. \cite{Takahashi2012} use the framework of the original \halofit of \cite{Smith2003}, but obtain improved accuracy by fitting to modern simulation data with superior resolution, extending to $k=30\iMpc$. The authors also focus their attention on models close to the current \LCDM paradigm, rather than more general models (such as those with power-law spectra, or curved models). \cite{Takahashi2012} used simulations of 16 different cosmological models around the best fits from the \textit{Wilkinson Microwave Anisotropy Probe} (\wmap) satellite (\wmap7 -- \citealt{Komatsu2011}; \wmap9 -- \citealt{Hinshaw2013}) and include models with $w\neq -1$. One can see how well \cite{Takahashi2012} compare to \emu in the lower row of Fig.~\ref{fig:best_fit} where \halofit can be seen to be comparable to our halo model but there is more high-$k$ spread at $z=0$ and a systematic over-prediction of the power around $k=1\iMpc$ that worsens with increasing redshift. \new{The stated accuracy of this version of \halofit is $5$ per cent for $k<1\iMpc$ and $10$ per cent up to $10\iMpc$, which is consistent with what is seen here.} A similar plot for the original \cite{Smith2003} version of \halofit shows large under-predictions for $k>0.5\iMpc$. From this point onwards we only compare to the revised \cite{Takahashi2012} version of \halofit.

\begin{table}
\caption{Cosmological parameters inferred from various data analyses. In all cases, we quote the best fit with $w=-1$ and flatness enforced.}
\begin{center}
\begin{tabular}{c c c c c c}
\hline
Cosmology & $\Ob$ & $\Om$ & $n_\mathrm{s}$ & $h$ & $\sigma_8$ \\
\hline
\wmap7 & 0.0457 & 0.275 & 0.969 & 0.701 & 0.810 \\
\wmap9 & 0.0473 & 0.291 & 0.969 & 0.690 & 0.826 \\
CFHTLenS & 0.0437 & 0.255 & 0.967 & 0.717 & 0.794 \\
\planck EE & 0.0487 & 0.286 & 0.973 & 0.702 & 0.796 \\
\planck All & 0.0492 & 0.314 & 0.965 & 0.673 & 0.831 \\
\hline
\end{tabular}
\end{center}
\label{tab:common}
\end{table}

\begin{figure}
\begin{center}
\includegraphics[width=6cm,angle=270]{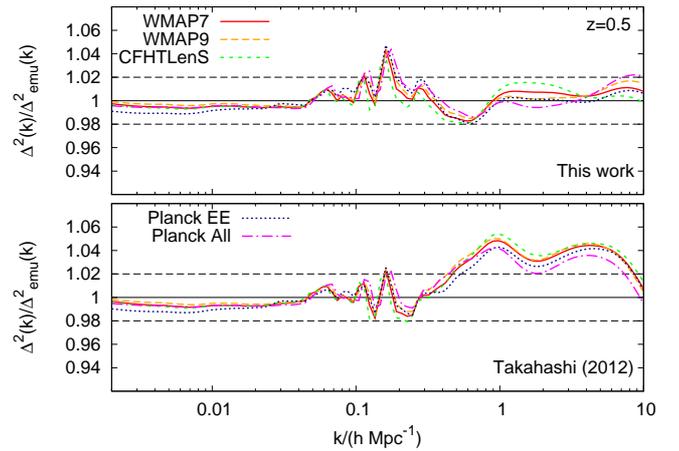}
\end{center}
\caption{A comparison of the power spectrum at $z=0.5$ of \meadfit and \halofit to that of \emu for several commonly used cosmological models (see Table~\ref{tab:common}) that derive from recent data sets. The error for each model is very similar because the cosmological models are all relatively similar. The \meadfit error rarely exceeds $2$ per cent, with the exception being around the BAO peak, which stems from our not modelling the non-linear damping of the BAO. The \halofit error rises to around $4$ per cent for $k>1\iMpc$ for all models.}
\label{fig:common}
\end{figure}

In Fig. \ref{fig:common} we show how our model fares for cosmological parameters derived from recent data sets (see Table~\ref{tab:common}). Once again we compare to \emu and show results for both our calibrated halo model and for the \cite{Takahashi2012} \halofit at $z=0.5$. One can see that the error from the halo-model approach rarely exceeds $2$ per cent for $k<10\iMpc$ for these cosmologies, with the worst error being an over prediction of the amplitude of the BAO peaks around $k=0.2\iMpc$. This arises because we did not attempt to model the exact non-linear damping of this feature in the power spectrum, and so our prediction here is very close to the undamped linear prediction. That our errors are better here than for the more general models shown in Fig.~\ref{fig:best_fit} is because these models all lie close to the centre of the \emu parameter space (see Table~\ref{tab:emu_parameters}). The \cite{Takahashi2012} \halofit model works better at BAO scales, but over-predicts the power at $k>0.5\iMpc$ systematically at around the $4$ per cent level.

\begin{figure}
\begin{center}
\includegraphics[width=6cm,angle=270]{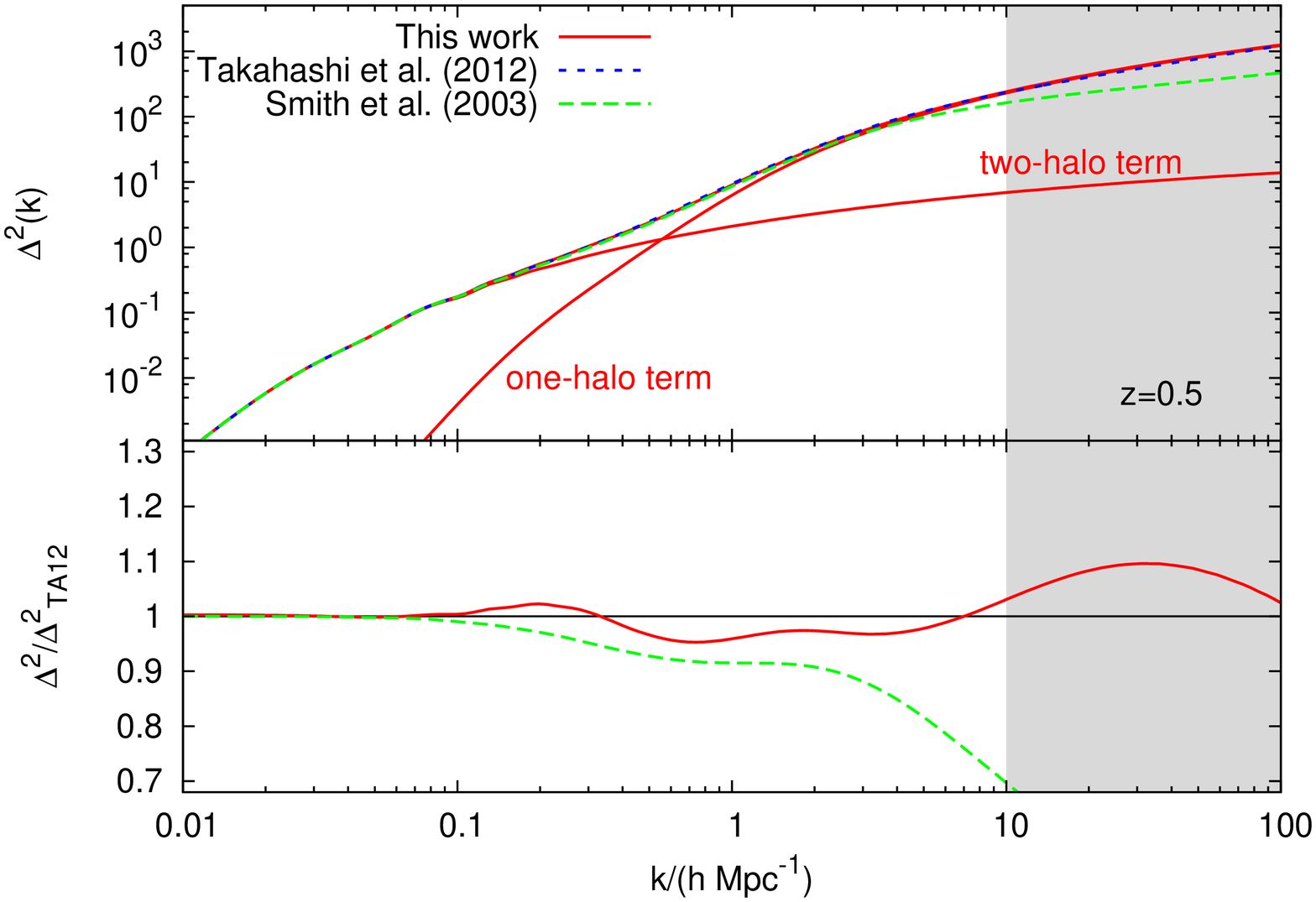}
\end{center}
\caption{The predictions of \meadfit compared to the two commonly used \halofit schemes up to $k=100\iMpc$ for a \planck cosmology at $z=0.5$. The upper panel shows $\Delta^2(k)$, while the lower panel shows the ratio of \meadfit to \citeauthor{Takahashi2012} (\citeyear{Takahashi2012}); we cannot show a comparison with \emu because it makes no predictions beyond $k=10\iMpc$. The range outside the bounds of \emu is marked by the grey region. The \meadfit and \citeauthor{Takahashi2012} (\citeyear{Takahashi2012}) models make predictions within $5$ per cent to $k=10\iMpc$ but this discrepancy increases to $10$ per cent for $k<100\iMpc$. Note that the power at these small scales is certainly strongly influenced by baryonic physics. The general level of agreement between these two models for a range of cosmologies can be inferred from Fig.~\ref{fig:best_fit}.}
\label{fig:high_k_power}
\end{figure}

The model presented here performs similar to, but slightly better than, the \cite{Takahashi2012} version of \halofit and has several advantages. Foremost, because we retain the apparatus of the halo model in our calculation, it means we can produce $\Delta^2(k)$ to arbitrarily high $k$ in a physically motivated way. Even though such extreme scales receive a small weight in lensing, they can be important if the modelling is badly wrong in this regime (\eg \citealt{Harnois-Deraps2015a}). A polynomial-based fitting formula such as \halofit risks generating pathological results when moving beyond the regime constrained by simulations and it is not at all obvious how to extend \emu. In Fig.~\ref{fig:high_k_power}, we show a comparison of the power spectrum predicted out to $k=100\iMpc$ with different models, simply to illustrate the range of behaviour at $k>10\iMpc$. Given that no simulations exist that could claim to accurately predict the matter power spectrum to $k=100\iMpc$ at $z=0$ we cannot make any quantitative statements about the accuracy of either model at these extreme scales, although both perform comparably. The grey shaded region in Fig.~\ref{fig:high_k_power} delimits these extreme scales and it is interesting to note that the maximum deviation between our model and \halofit is only $\sim 10$ per cent.

Recently an approach related to ours has been pursued by \cite{Mohammed2014a} and \cite{Seljak2015}; these authors use a combination of perturbation theory for a two-halo term and a polynomial series expansion for a one-halo term, constrained to contain only even powers of $k$ by theoretical considerations. These authors obtain remarkable fits to the matter power spectrum and correlation function, but at the cost of fitting each term in the one-halo series expansion to simulations up to $k\simeq 1\iMpc$. In \cite{Mohammed2014a}, it was shown that fits accurate at the $2$ per cent level were possible to the \emu nodes up to $k=0.3\iMpc$ but their fit degrades at smaller scales; it was tested out to $0.7\iMpc$ and it is not obvious how to extend predictions to smaller scales. Our approach utilizes the full apparatus of the halo model and can therefore be extrapolated with a degree of robustness to the smaller scales that are required to make predictions for weak-lensing observables, which we show in section~\ref{sec:lensing}. We are also in a position to be able to model the effects of baryonic feedback in a physically motivated way, and we now turn the attention of the reader to this subject.

\section{Baryonic physics}
\label{sec:baryons}

Whilst $5$ per cent accuracy across a range of cosmologies and scales is an important achievement of this work, it is clear that baryonic processes can have a much larger impact on the non-linear mass distribution than incorrect modelling of the dark-matter only spectrum (\eg \citealt{White2004}; \new{\citealt{Zhan2004}}; \citealt{Zentner2008}; \new{\citealt{Casarini2011}}; \citealt{Zentner2013}; \citealt{Semboloni2013}; \citealt{Eifler2014}; \citealt{Harnois-Deraps2015a}). Baryonic physics is not normally accounted for in numerical simulations and baryons can undergo processes such as radiative cooling where they collect in sufficient density. The gas then contracts, which alters the dark matter distribution via gravitational interactions, and so the total matter distribution is altered because neither the baryons nor dark matter are where they would be if only gravitational interactions were considered (\eg \new{ \citealt{Jing2006}}; \citealt{Duffy2010}). Alternatively, supernova explosions or energy released by active galactic nuclei (AGN) can heat gas, which can then expand outside of the virial radius of haloes (\citealt{Schaye2010}; \citealt{vanDaalen2011}; \new{ \citealt{Martizzi2014}}; \citealt{vanDaalen2015}), and as a consequence the total matter distribution within a halo can be altered significantly. AGN feedback can reduce the baryon fraction in the centres of haloes by a factor of $2$ in the most extreme models \citep{Duffy2010}.

An advantage of the approach advocated in this work is that one might attempt to capture the influence of baryons on the matter power spectrum simply by varying parameters that control the internal structure of haloes. This is possible because we retain the theoretical halo-model apparatus in \meadfit. Physically, one can regard baryonic processes as altering the internal structure of haloes, while not affecting their positions or masses to the same degree (\eg \citealt{vanDaalen2014}). It has been demonstrated that the effect of baryons can be captured by altering the halo internal structure relations, using information that is measured in baryonic simulations (\citealt{Zentner2008}; \citealt{Duffy2010}; \citealt{Zentner2013}; \citealt{Semboloni2013}; \citealt{Mohammed2014b}). The general trend is that gas cooling increases the central density of haloes whereas violent feedback, such as that from AGN, decreases the concentration. How this translates into the matter power spectrum in simulations is considered in \cite{vanDaalen2011} where it was shown that per cent level changes in $\Delta^2(k)$ can arise at $k=0.3\iMpc$ as a result of strong AGN feedback. \cite{Semboloni2011}, \cite{Eifler2014} and \cite{Mohammed2014b} all showed that failing to account for feedback would strongly bias cosmological constraints from the weak lensing Dark Energy Survey if the most extreme feedback scenarios apply to our Universe and constraints from \euclid would be severely biased for any feasible feedback scenario. \cite{Duffy2010}, \cite{Semboloni2011} and \cite{Mohammed2014b} also showed that the main effects of baryonic feedback could be captured using a halo-model prescription, considering how feedback would alter the internal structure of haloes.

\begin{figure}
\begin{center}
\includegraphics[width=9cm,angle=270]{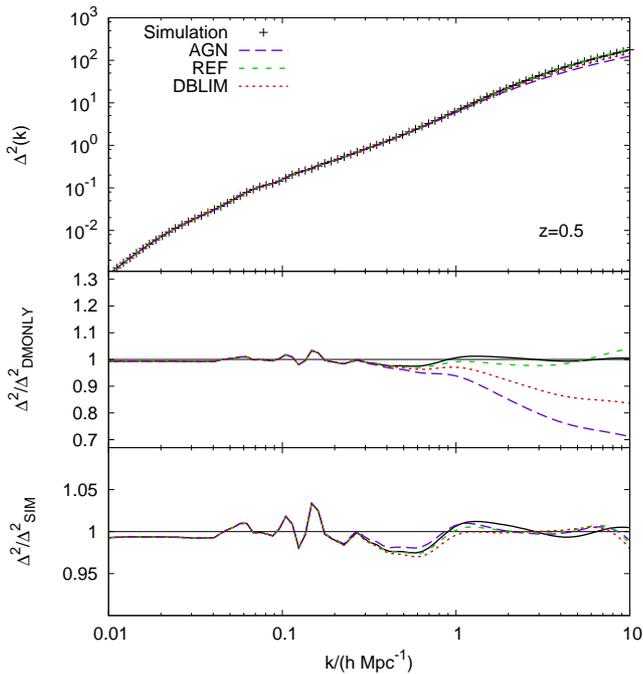}
\end{center}
\caption{\new{Best fitting halo-model power to the power spectra of the OWLS simulations for the \dmonly (black; solid), \agn (purple; long-dashed), \refe (green; medium-dashed) and \dblim (red; short-dashed) models up to $k=10\iMpc$ at $z=0.5$. These are obtained by fitting both $A$ and $\eta_0$ (equations \ref{eq:bullock_cm} and \ref{eq:eta_definition}) to each model at this redshift. In the top panel, we show $\Delta^2$ while in the middle panel we show the ratio of each spectrum to the emulator \dmonly case (black crosses in the top panel); one can see that the freedom introduced by allowing these parameters to vary is able to capture both the up- and down-turn in power that feedback introduces relative to the dark-matter-only case. Any residual differences for $k\ltsim1\iMpc$ are due to residual errors in our fitting across a range of cosmologies that can be seen in Fig. \ref{fig:best_fit}. Our accuracy is best appreciated in the lower panel, in which we show the ratio of each halo-model prediction to the corresponding simulation.}}
\label{fig:baryon_power}
\end{figure}

We use power spectra from the OverWhelmingly Large Simulations (OWLS; \citealt{Schaye2010}; spectra from \citealt{vanDaalen2011}) of a dark-matter (\dmonly) model; a model that has prescriptions for gas cooling, heating, star formation and evolution, chemical enrichment and supernovae feedback (\refe); a model that is similar to \refe but with the addition of active galactic nuclei (AGN) feedback (called \agn); and a model similar to \refe but which additionally has a top-heavy stellar initial mass function and extra supernova energy in wind velocity (\dblim -- called \textsc{dblimfv1618} in \citealt{vanDaalen2011}). It was shown in \cite{vanDaalen2011} that the difference in power between the \dmonly and \agn models is particularly large.

We fit the power spectra from the OWLS simulations using our calibrated halo-model approach with a halo profile that is altered to reflect baryon bloating and gas cooling. Again, our new fitted halo profiles may not match those of simulated haloes exactly but our aim is to match the \emph{power spectrum} accurately. However, we would expect the trends observed in the profiles of simulated haloes to be respected by any modification to halo profiles in \meadfit. For example, if we require an enhanced concentration to fit data for a particular model in OWLS, then haloes measured in this baryonic model should display enhanced concentrations relative to their \dmonly counterparts. This approach differs from that presented in \cite{Semboloni2013}, \cite{Fedeli2014a}, \cite{Fedeli2014b} and \cite{Mohammed2014b} in that we do not attempt to add accurate profiles for the gas and stars into the halo model, but instead look for a more empirical modification that is able to match data at the level of the power spectrum for $k<10\iMpc$.

\begin{figure}
\begin{center}
\includegraphics[width=6cm,angle=270]{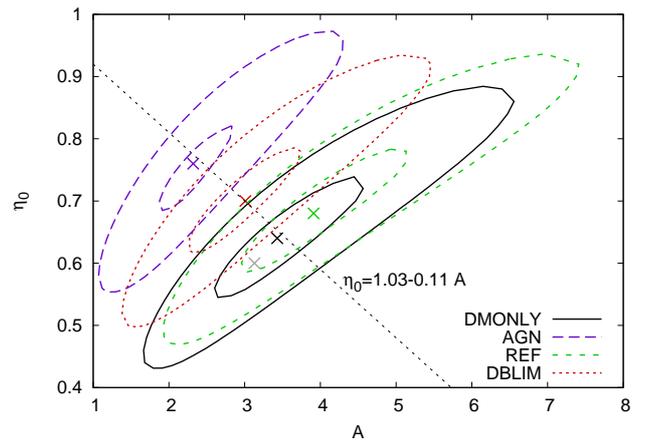}
\end{center}
\caption{Best matches to the power spectrum from the OWLS simulations found by varying halo structure via $A$ and $\eta_0$ (equations \ref{eq:bullock_cm}, \ref{eq:eta_definition} and \ref{eq:eta0_definition}) from $z=0$ to $1$. The contours enclose regions of parameter space that match the power spectra with an average error of $2.5$ per cent (inner) and $5$ per cent (outer) from $k=0.01$ to $10\iMpc$ and the crosses mark the best-fitting point. \new{We show contours for the \dmonly (black; solid), \agn (purple; long-dashed) \refe (green; medium-dashed) and \dblim (red; short-dashed) cases.} These ranges can be used to place a prior on the range of $\eta_0$ and $A$ to be explored in a cosmological analysis as they encompass the range of behaviour expected from plausible feedback models. The dashed line (equation~\ref{eq:one_param_fit}) shows a relation between $\eta_0$ and $A$ that could be used to provide a single-parameter fit to all models. The grey cross is the best-fitting value to all the \emu simulations, whereas the black cross is the best match to the specific cosmology used in the \dmonly model.}
\label{fig:contours}
\end{figure}

Given the above discussion, we might expect that two parameters would suffice: one to capture the increased concentration as gas cools in halo cores and one to capture the puffing up of halo profiles due to more violent feedback. To fit the baryonic models, we allow ourselves to vary the parameter $A$ in the $c(M)$ relation (equation \ref{eq:bullock_cm}) and the parameter $\eta_0$, where $\eta$ is defined in equation~(\ref{eq:eta_definition}) and $\eta_0$ is the first parameter in the full expression for $\eta$ in Table~\ref{tab:fit_params}, explicitly
\begin{equation}
\eta=\eta_0-0.3\,\sigma_8(z)\ .
\label{eq:eta0_definition}
\end{equation}
All other parameters are fixed to their values in Table~\ref{tab:fit_params}, and the redshift evolution in the second half of the expression for $\eta$ is preserved from the best fit to the \emu nodes. We vary $\eta_0$ and $A$ to best fit the OWLS data from simulations \dmonly, \agn, \refe and \dblim. We construct these power spectra by taking ratios of the publicly available OWLS baryon models to the \dmonly models (which produces a smooth curve because the simulations have matched initial conditions) and then multiplying this ratio by the \emu prediction for the baseline \wmap3 cosmology used for the OWLS simulations. We do this because the OWLS simulations are small in volume and the power spectrum would be too noisy to use in its raw form. The best fits to OWLS power spectra are shown in Fig.~\ref{fig:baryon_power} where it can be observed that the freedom permitted by fitting $A$ and $\eta_0$ allows the power spectrum of \meadfit to trace the residual displayed by the OWLS simulations accurately over the range of scales shown. Particularly, note that the variation is able to reproduce both the up-turn due to gas cooling, enhancing clustering around $k=10\iMpc$, and the down-turn due to mass being expelled from the halo, which can impact the relatively large scale of $k=0.3\iMpc$.

In Fig.~\ref{fig:contours} we show how the goodness-of-fit varies as parameters $A$ and $\eta$ are varied for the various feedback recipes. The contours enclose regions of parameter space in which the average error is $2.5$ per cent (inner) and $5$ per cent (outer), where the average is taken over all scales between $k=0.01$ and $10\iMpc$, binned logarithmically. One can see that these parameters distinguish well between the simulated \agn model and the other models, \dblim is marginally distinguished, but parameters that fitted \dmonly and \refe best are nearly identical. The distinguishability is directly related to the magnitude of the effect that each model has on the power spectrum (for $k<10\iMpc$), which can be seen in the middle panel of Fig.~\ref{fig:baryon_power}. Our best-fitting parameters for each model are given in Table~\ref{tab:baryon_params}. The \agn model clearly favours less concentrated haloes, which is expected given that AGN blow gas out of the central portions of haloes. The range of acceptable parameter combinations of $A$ and $\eta_0$ that are able to fit the OWLS data, shown in Fig.~\ref{fig:contours} could be used to form a prior in future weak-lensing analyses that aim to constrain or marginalize over baryonic feedback. This is acceptable because it is not clear which, if any, of the OWLS models is the correct one. The most conservative assumption is that the recipes give a range of plausible feedback effects and this is what the range we suggest in Fig.~\ref{fig:contours} encapsulates. Additionally, the dashed line shows a relation between $\eta_0$ and $A$ that could be used to provide a \emph{single} parameter fit to all models:
\begin{equation}
\eta_0=1.03-0.11A\ .
\label{eq:one_param_fit}
\end{equation}
This relation exists because there is some degeneracy between the effects of varying $\eta$ and those of varying $A$. Applying this relation would mean that the \refe model could not be distinguished from \dmonly, but all other models could be distinguished. A further advantage of the halo-model approach is that it should also capture any coupling between cosmological parameter variation and feedback processes. This effect is ignored in any polynomial- or template-based approach to modelling feedback (\eg \citealt{Eifler2014}). 

\begin{table}
\caption{Parameter combinations of $\eta_0$ and $A$ that best fit OWLS data from $z=0$ to $1$ via the halo-model approach described in the text. These parameters are those at the centres of the ellipses in Fig.~\ref{fig:contours}. The OWLS simulations can be matched at the $5$ per cent level over the redshift range. That the values of $\eta_0$ and $A$ differ in the case of `all \emu simulations' compared to \dmonly is because a slightly improved fit is possible in the case of dealing with a specific cosmology, which in the case of OWLS is the slightly outdated \wmap3 ($\Om=0.238$, $\Ob=0.0418$, $\sigma_8=0.74$, $n_\mathrm{s}=0.951$, $h=0.73$).}
\begin{center}
\begin{tabular}{c c c}
\hline
Model & $\eta_0$ & $A$ \\
\hline
All \emu simulations & 0.60 & 3.13 \\
\dmonly (\wmap3 from OWLS) & 0.64 & 3.43\\
\agn & 0.76 & 2.32\\
\refe & 0.68 & 3.91\\
\dblim & 0.70 & 3.01\\
\hline
\end{tabular}
\end{center}
\label{tab:baryon_params}
\end{table}

\section{Weak gravitational lensing}
\label{sec:lensing}

The origin of the late-time accelerated expansion of the cosmos is uncertain. To test different models it is necessary to measure the expansion rate together with the growth of perturbations around the present day. The weak gravitational lensing of galaxies has emerged as one of the premier tools to probe perturbations, which can discriminate between models with similar background expansion rates.

However, the matter power spectrum is not a directly measurable quantity. With the notable exception of \cite{Brown2003} and the 3D lensing of \cite{Kitching2014}, the majority of weak-lensing analyses have measured real space angular correlations of the shears of galaxy images. Shear is typically expressed in terms of the spin-2 quantity $\gamma$, which can be split into tangential ($\gamma_\mathrm{t}$) and cross ($\gamma_\times$) components with respect to a pair separation vector. Correlation functions $\xi_\pm$ are defined as combinations of two-point correlations of tangential- and cross-shear as a function of angular separation:
\begin{equation}
\xi_\pm=\average{\gamma_\mathrm{t}\gamma_\mathrm{t}}\pm\average{{\gamma_\times}{\gamma_\times}}\ .
\end{equation}
These can be related to the harmonic coefficients, $C(\ell)$, of the spherical Fourier transform of a weighted, projected matter field \citep{Kaiser1992},
\begin{equation}
\xi_\pm(\theta)=\frac{1}{2\pi}\int_0^\infty C(\ell)J_\pm(\ell\theta)\ell\;\mathrm{d}\ell\ ,
\label{eq:lensing_xi}
\end{equation}
with harmonic wavenumber $\ell$. Here $J_\pm$ is the zeroth ($\xi_+$) and fourth ($\xi_-$) Bessel functions respectively. In the \cite{Limber1954} approximation the $C(\ell)$ are then related to integrals over all scales of the matter power spectrum \citep{Kaiser1992}:
\begin{equation}
C(\ell)=\int_0^{\omega_\mathrm{H}} \frac{g^2(\omega)}{a^2(\omega)}P(k=\ell/f_K(\omega),z(\omega))\;\mathrm{d}\omega\ ,
\label{eq:lensing_cl}
\end{equation}
where $\omega(z)$ is the comoving distance to redshift $z$, defined via the metric convention of \cite{Bartelmann2001} and $g(\omega)$ is the lensing efficiency function:
\begin{equation}
g(\omega)=\frac{3H_0^2\Om}{2c^2}\int_\omega^{\omega_\mathrm{H}}n(\omega')\frac{f_K(\omega'-\omega)}{f_K(\omega')}\;\mathrm{d}\omega'\ .
\end{equation}
Here $f_K(\omega)$ is the comoving angular-diameter distance, $\omega_\mathrm{H}$ is the comoving distance to the horizon and $n(\omega)$ is the normalized distribution of source galaxies. Finally, $P(k)$ is the power spectrum of matter fluctuations, defined in equation~(\ref{eq:P(k)}), and is exactly what \meadfit provides. Thus, to make contact with lensing observables, it is necessary to investigate how the accuracy of the predictions of \meadfit at the level of the matter power spectrum translates into accuracy for $\xi_\pm$.

\begin{figure}
\begin{center}
\includegraphics[width=5.9cm,angle=270]{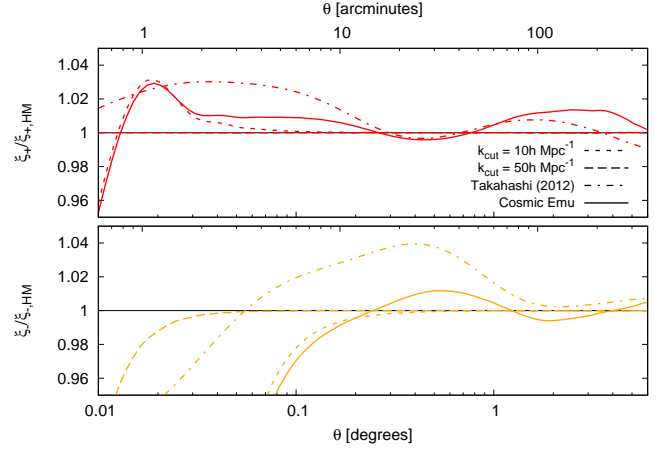}
\end{center}
\caption{Lensing correlation functions predicted by various methods are shown divided by the correlation functions predicted using \meadfit with no $k$-cut imposed. $\xi_+$ ( red) is shown in the upper panel and $\xi_-$ (orange) in the lower panel. Source galaxies are all taken to be fixed at $z_\mathrm{s}=0.7$, the effective median lensing redshift of \cfhtlens. We show the cases of taking the \meadfit matter power spectrum to be sharply cut at $k=10h$ \new{(short-dashed)} and $50\iMpc$ \new{(long-dashed)} and correlation functions from \halofit \new{ (dot-dashed)} and \emu (\new{solid}; where $\Delta^2=0$ for $k>10\iMpc$). For the range of angular scales shown, all models agree for $\xi_+$ at the $\simeq3$ per cent level, a $k$-cut at $k=10\iMpc$ only starts to impact upon $\xi_+$ at the per cent level for $\theta < 0.03^\circ$ and a cut at $50\iMpc$ has almost no impact for angles shown; the agreement between \meadfit and \emu is $\simeq2$ per cent across the range of angles. For $\xi_-$, \meadfit and \emu agree to $\simeq 2$ per cent until the $k$-cut becomes important, around $\theta=0.2^\circ$. \halofit is discrepant at the $4$ per cent level for $\theta<0.5^\circ$. For $\xi_-$, the impact of $k$-cuts is felt at larger angular scales than for $\xi_+$ -- a cut at $50\iMpc$ makes its impact at the per cent level around $\theta=0.02^\circ$.}
\label{fig:xi}
\end{figure}

Several methods are commonly used to deal with the fact that 2D lensing correlation functions depend on integrals over all $k$ of the matter spectrum. Either fitting formulae are extrapolated beyond the regime to which they were fitted, in the hope that they still provide the required accuracy, or $P(k)$ is set to zero for regions where the power is unknown, a so-called $k$-cut. As previously stated, one of the benefits of our halo-model approach is that we have more reason to trust the predictions of the halo model at small scales, due to the large amount of theoretical input that goes into the model. In Fig.~\ref{fig:xi} we show the lensing $\xi_\pm$ correlation functions as predicted by integrating over $P(k)$ from \meadfit, \halofit and from \emu (where we set $\Delta^2(k)=0$ for $k>10\iMpc$). These are computed by taking the source galaxies to all be fixed at $z_\mathrm{s}=0.7$, the effective median redshift for lensing of \cfhtlens \citep{Heymans2012}, and using the best-fitting cosmology from the \cite{PlanckXIII2015}. The three models for $\xi_+$ agree at the $\simeq 3$ per cent level for at all angular scales shown with predictions from \meadfit and \emu agreeing at the per cent level for all angles shown. We also show \meadfit $\xi_\pm$ predictions for $k$-cuts at $10$ and $50\iMpc$. If the theory were perfect, a cut-off in $\Delta^2$ at $k=10\iMpc$ provides per cent level accuracy only for $\theta>0.03^\circ$; if the power spectrum is cut at $50\iMpc$, then the predicted $\xi_+$ does not deviate from per cent level agreement for all angles shown. For $\xi_-$ if no $k$-cut is imposed, the three models agree at the per cent level only for $\theta>1^\circ$, after which predictions from \cite{Takahashi2012} deviate by as much as $6$ per cent at $\theta\simeq 0.01^\circ$. The effect of the finite resolution of \emu becomes important at the per cent level at $\theta=0.2^\circ$ and the same deviation can be seen in the \meadfit prediction when it is cut at $k=10\iMpc$. Assuming perfect knowledge of the theory this accuracy extends to $\theta>0.02^\circ$ if the cut is taken at $50\iMpc$. We note that if we extend the matter spectrum from \emu using a power law for $k>10\iMpc$ the predictions agree with those of \meadfit to $2$ per cent for both $\xi_\pm$; this is probably due to the fact that the halo-model prediction at $k>10\iMpc$ involves an integral over many quantities that are accurately power laws, resulting in a close to power-law power spectrum.

\begin{figure}
\begin{center}
\includegraphics[width=3.5cm,angle=270]{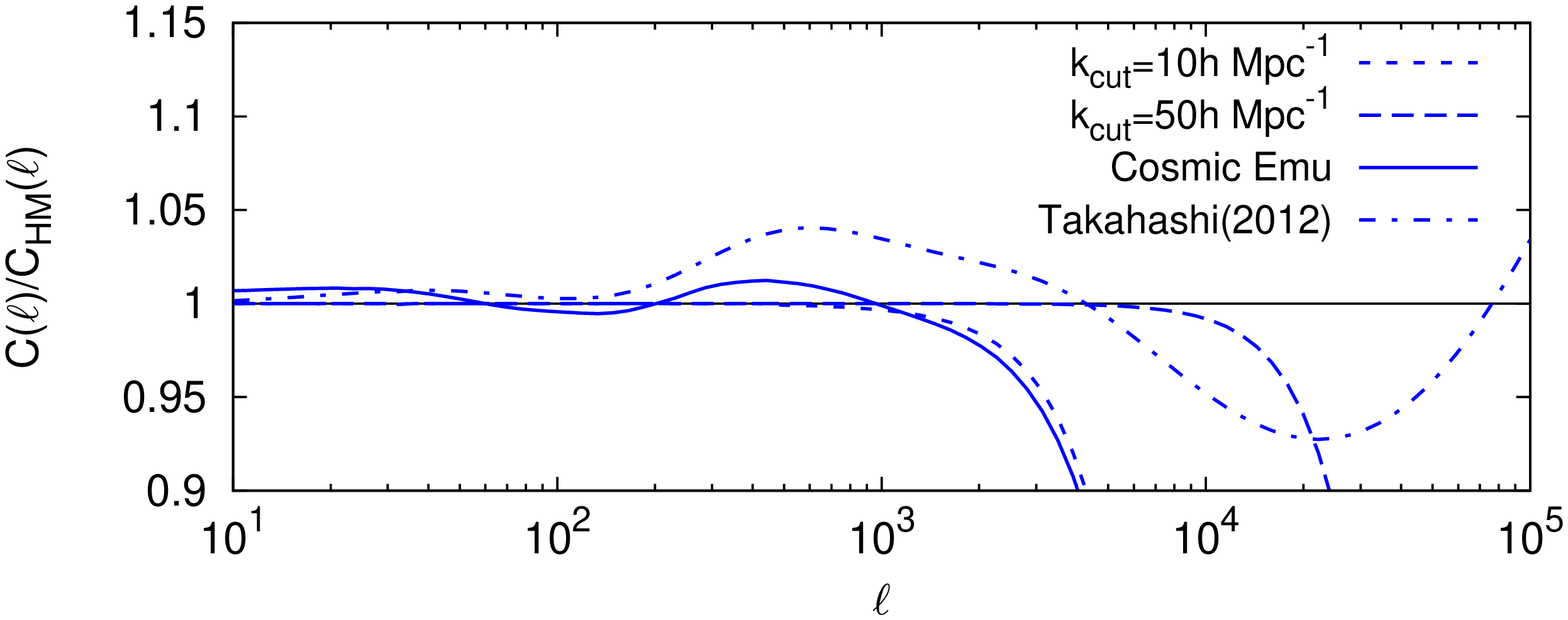}
\end{center}
\caption{As Fig. \ref{fig:xi} but for the $C(\ell)$ coefficients. Each model shown is divided by the \meadfit prediction with no $k$-cut imposed. We see that a cut in $\Delta^2$ at $k=10\iMpc$ induces per cent level deviations around $\ell=10^3$ and a cut at $50\iMpc$ increases this to $\ell=10^4$. \meadfit and \emu agree at the per cent level until the finite $k$-range of \emu becomes important. \halofit disagrees by as much as $5$ per cent around $\ell=10^3$ and this disagreement increases to $7$ per cent for higher harmonics.}
\label{fig:cl}
\end{figure}

In Fig. \ref{fig:cl} we compare models at the level of their $C(\ell)$ predictions. When the finite $k$-range of \emu is unimportant, it agrees with the \meadfit prediction to one per cent. Discrepancies with \halofit at the $5$ per cent level arise from $\ell=500$ with maximum deviations of $7$ per cent for $\ell>10^4$. A cut in power at $k=10\iMpc$ impacts upon the $C(\ell)$ at the per cent level around $\ell=10^3$ and a cut at $50\iMpc$ at $\ell=10^4$.

\begin{figure}
\begin{center}
\includegraphics[width=8.4cm,angle=270]{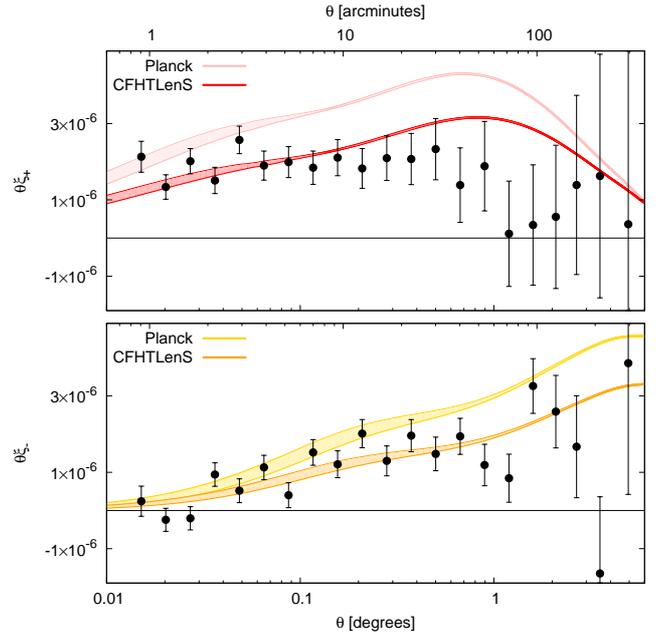}
\end{center}
\caption{$\xi_+$ (upper panel) and $\xi_-$ (lower panel) correlation functions predicted using \meadfit, with a width showing the spread that is obtained from different feedback models. In each case, source galaxies are taken to be fixed at $z_\mathrm{s}=0.7$, approximately the effective median redshift for lensing for \cfhtlens, and we show the correlation functions predicted using the best fitting \textit{Planck} cosmology (upper curve) and the best fitting \cfhtlens \citep{Heymans2013} cosmology (lower curve). For comparison, we also show the measured $\xi_\pm$ from the \cfhtlens survey; it can be seen that feedback fails to alleviate the tension between \cfhtlens and \textit{Planck} data. One can see that an ignorance of the details of feedback affects $\xi_-$ to much larger angular scales than $\xi_+$, a consequence of it probing more non-linear regions of the matter distribution. Baryonic feedback has an impact at the greater than per cent level for $\theta<0.1^\circ$ for $\xi_+$ and $\theta<2^\circ$ for $\xi_-$. In all cases, the effects of baryonic feedback are small relative to the errors in current data.}
\label{fig:baryon_correlation}
\end{figure}

\new{Many authors have investigated how baryonic processes affect weak-lensing observables. Early work (\citealt{White2004}; \citealt{Zhan2004}) used the halo model to estimate how the matter power would be altered by including gas cooling in haloes and hot, diffuse intra-cluster gas, respectively. More recent work has used hydrodynamic simulations with feedback recipes (\eg \citealt{Jing2006}; \citealt{Semboloni2011}; \citealt{Casarini2012}) to compare weak-lensing observables to the case when no baryonic feedback is included. The results are that for $\ell>1000$, the $C(\ell)$ are altered at the per cent level with the alterations increasing with $\ell$, but the details are strongly dependent on the feedback implementation.}

In Fig.~\ref{fig:baryon_correlation} we show the range of possible correlation function predictions given by our power spectrum fits to the baryonic feedback models, where the region enclosed by the curves is the region that our fits to the OWLS feedback models occupy (the centres of the ellipses in Fig.~\ref{fig:contours}). We generate these using \meadfit predictions with parameters $A$ and $\eta_0$ taken from the centres of the ellipses in Fig.~\ref{fig:contours}. We also show data from the \cfhtlens analysis of \cite{Kilbinger2013} so that ignorance of the details of feedback can be compared to the current errors in data. For current data we see that the effect of feedback is small compared to the errors, but data that will be available in the near future will increase in accuracy and feedback processes will have to be accounted for. Fig.~\ref{fig:baryon_correlation} also shows that baryonic feedback does little to alleviate the tension between the best-fitting \cfhtlens cosmology and that of \textit{Planck}. For our case of sources fixed at $z_\mathrm{s}=0.7$ baryonic feedback only has an effect at the greater than per cent level for $\theta<0.1^\circ$ for $\xi_+$ and $\theta<1^\circ$ for $\xi_-$. In a forthcoming paper (Joudaki et al. in prep) constraints on $A$ and $\eta_0$, together with cosmological constraints when these parameters are marginalized over, will be presented using the \cfhtlens together with that from RCSLenS. 

Alternative approaches have been investigated to model the impact of feedback on weak-lensing observables, all of which use data from the OWLS hydrodynamic simulations: \cite{Mohammed2014a} model the OWLS data by refitting the coefficients from their power-series expansion of the one-halo term and advocate marginalizing over these coefficients to immunize against biases due to feedback. \cite{Harnois-Deraps2015a} construct polynomial fits to the ratio of power spectra from feedback models to the \dmonly model; again the coefficients of these polynomials could feasibly be constrained by data. However, to fit each model over the scale redshift range required 15 coefficients, compared to only 2 in our approach. With a fixed (\wmap9) cosmology, \cite{Harnois-Deraps2015a} find a preference for feedback in the \cfhtlens data. \cite{MacCrann2015} reanalysed the \cfhtlens survey but adding a single parameter that governs the amplitude of AGN feedback, which was taken to be given by the ratio of power from the \agn to \dmonly simulations. They find only a weak preference for feedback, but find that AGN feedback is insufficient to resolve tension between the \cfhtlens data and that of the \textit{Planck} satellite. An identical conclusion is reached by \cite{Kitching2014} using a similar method. \cite{Eifler2014} propose using a principal component analysis method, by which components of the power spectrum that are most affected by feedback are removed. Assuming that feedback is independent of cosmology, they show that the data could be fitted with as few as 4 components removed. Our method may be preferable to this as it potentially allows one to capture the coupling between feedback and cosmological parameters, via their effects on the halo profiles.

\section{Summary and discussion}
\label{sec:conclusion}

We have shown that the halo model can be optimized so that it accurately reproduces power spectra measured from dark-matter \nbody simulations across a range of cosmological models for $k\leq10\iMpc$ and $z\leq2$, provided we are willing to introduce a number of empirical modifications to its ingredients. We achieved this by calibrating our model to the `node' simulations of \emu of \cite{Heitmann2014}. Our success reflects the fact that the halo model is built on well-posed theoretical ingredients, which naturally adapt to changes in cosmology in a sensible fashion. Our fits are accurate at the $5$ per cent level, which represents an improvement over the currently used \halofit model of \cite{Takahashi2012}. \emu itself is quoted to be accurate to $5$ per cent at $k=10\iMpc$ and so our fit is as good as possible at the current level of ignorance at this scale. Even in the dark-matter-only case, our accuracy statement comes with the caveat that we have only tested a limited range of plausibly interesting cosmologies. In particular, we have concentrated on the parameter cube of the \emu of \cite{Heitmann2014}, which contains cosmological parameters within the $5\sigma$ region of the \cite{PlanckXIII2015} results for the standard cosmological model.

The advantage of the halo-model approach is that it can be simply expanded beyond the parameter cube of \emu. We demonstrated that our model is able to produce reasonable power for $k>10\iMpc$ and it can also produce spectra for $z>4$, greater than allowed by \emu. Given the large amount of tested theoretical input that goes into the halo-model calculation, we expect that our model should produce sensible spectra for higher redshifts and smaller scales. For small deviations from the standard cosmological paradigm, such as dark energy with time-varying equation of state, or for models with small amounts of curvature, we also expect our answers to be accurate. If one were interested in the power spectrum of radically different models, such as very curved models or those qualitatively different linear power-spectrum shapes, we would advise caution, as our model has not been tested to these extremes. 

\new{Our approach differs from that of some authors (\eg \citealt{Cooray2001}; \citealt{Giocoli2010}) who attempt to improve the basic halo model by adding effects such as halo substructure and a scatter in halo properties at a given mass. It may be that adding in these ingredients would have reduced the number or magnitude of our fitted parameters, but we make no attempt to quantify this.} Our approach also contrasts with that employed by \cite{Mohammed2014a} and \cite{Seljak2015} who advocate replacing a theoretically motivated one-halo term with a power-series expansion and fitting this expansion term-by-term. Combining this approach with perturbation theory produced excellent results in the quasi-linear ($k\simeq 0.3\iMpc$) regime. However, it is not obvious how to extend their predictions to smaller scales than the smallest constrained by these authors ($k\simeq 1\iMpc$), as their empirical power series has no physical requirement for sensible behaviour at smaller scales. Given the necessity of these smaller scales in producing usable weak-lensing predictions, we therefore prefer our approach. A future avenue of fruitful research may be to combine the power of our approach of a physically motivated fit to the one-halo term with the perturbation-theory-inspired two-halo term advocated in these works. Particularly this may improve accuracy in the quasi-linear regime and would certainly help in modelling around the BAO scale, where our $5$ per cent accuracy deviates from the per cent level accuracy of \emu.

Baryonic feedback has a large impact on the power at small scales. We demonstrated that the halo model is able to capture the influence of baryonic physics via only two redshift-independent parameters that govern halo internal structure. We also showed that this can be reduced to one parameter, at the loss of a small amount of discriminating power. Using these two parameters, we were able to fit feedback recipes considered in the OWLS simulations at the 5 per cent level for $z\leq1$ and $k\leq10\iMpc$. Because these parameters are firmly rooted in the halo-model apparatus, their effects are restricted to scales at which haloes affect the power spectrum. This is not guaranteed by other approaches, such as polynomial fits, which may produce unphysical effects. We also suggested that our approach is more likely to reproduce the correct \emph{coupling} between baryonic feedback and cosmology, because it is rooted in halo properties. It is not obvious how to account for the cosmology dependence of baryonic feedback using existing fitting formulae or \emu. 

Finally, we showed how the power-spectrum predictions of \meadfit translate into the lensing $C(\ell)$ and $\xi_\pm$ correlation functions that are measured in a standard lensing analysis. For the dark-matter-only case, we showed that \meadfit agrees with \emu at the per cent level for $\xi_+$  and $C(\ell)$ and $2$ per cent for $\xi_-$ for all scales where the finite $k$-range of \emu is unimportant. We suggest that \meadfit provides a reasonable way of extrapolating \emu to the smaller scales that are necessary to produce weak-lensing predictions. Considering baryonic feedback, we showed how our matter power spectra translate into lensing predictions and showed that ignorance of the details of feedback is smaller than uncertainty on current data. For a survey similar to \cfhtlens, baryonic feedback impacts on $\xi_+$ at the per cent level for $\theta<0.1^\circ$ and on $\xi_-$ for $\theta<2^\circ$. In future lensing analyses (\eg Joudaki et al., in preparation) we advocate marginalising over the halo parameters that we used to fit to the OWLS feedback models in order to produce unbiased cosmological constraints; our range of fitted values for these parameters may be used as a prior for this purpose. Alternatively, one might accept a given cosmology and then use the best-fitting baryon parameters as a means of learning about baryonic feedback.

In summary, and given our accuracy, we suggest that \emu be used if one is interested in the non-linear, gravity-only induced power spectrum for $k\leq10\iMpc$, and the required model is within the \emu parameter cube. However, if one is interested in departures from the \emu parameter space, accounting for the effect of baryonic feedback physics, or producing accurate lensing observables via a reasonable extrapolation, we then advocate \meadfit. Although we focused on weak lensing in this paper, we stress that \meadfit is useful for any application that currently uses \halofit.

This last point emphasizes the potential of the approach described in this paper. The halo model can readily be extended to take account of new physical processes and changes in the cosmological paradigm.  One example for further work would be an application of our method to cover modified gravity models (\eg \citealt{Schmidt2010a}; \citealt{Lombriser2014a}; \citealt{Barreira2014}) where revised growth rates, collapse thresholds and internal halo structures can be predicted in part on analytic grounds, and where there is a growing effort on detailed simulations. Another such example would be to look at the impact of massive neutrinos (\eg \citealt{Massara2014}) on the matter spectrum. In such cases, being able to produce \emph{accurate} power spectra will be important in order to distinguish standard and non-standard cosmological models. Moreover, exploration of a large parameter space of models will inevitably be necessary, and there will therefore be a strong motivation to explore rapid means of generating non-linear power spectra.

The code developed as part of this paper is available at \meadaddress or at request from the author. \new{It is able to produce the matter power spectrum at 200 $k$ values for $16$ different $z$ values in $\simeq 0.5\mathrm{s}$ on a single core of a standard desktop computer.}

\section*{Acknowledgements}

AJM acknowledges the support of an STFC studentship and, together with CH, support from the European Research Council under the EC FP7 grant number 240185. Part of this work was carried out at the 2014 `Modern Cosmology' conference at the Centro de Ciencias in Benasque, Spain. \new{We also acknowledge an anonymous referee for useful comments.}

\label{lastpage}

\footnotesize{
\setlength{\bibhang}{2.0em}
\setlength\labelwidth{0.0em}
\bibliographystyle{mnras}
\bibliography{meadbib}
}

\appendix

\section{Power spectrum response}
\label{app:response}

\begin{figure*}
\begin{center}
\includegraphics[width=17.5cm]{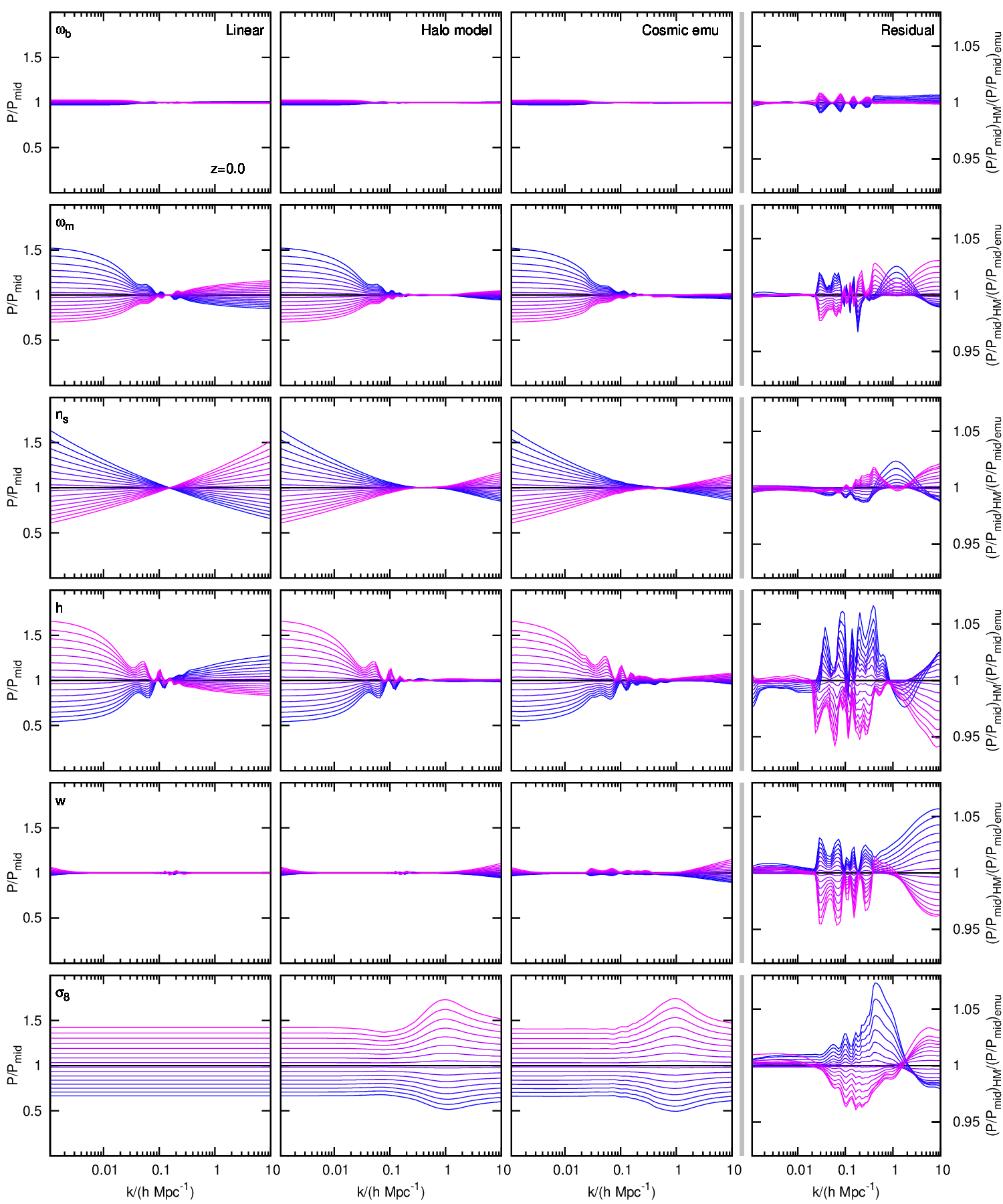}
\end{center}
\caption{The ratio of matter power at $z=0$ as cosmological parameters are varied when compared to a `fiducial' cosmological model of the \emu parameter space. We show comparisons of linear theory (left column), \meadfit (second column) and the \emu prediction (third column). Comparing the two central columns, one can see that in general \meadfit is able to reproduce the trends in power-spectrum response to cosmological parameter variation accurately. The right-hand column shows a residual of the \meadfit power divided by that of the emulator \newtwo{(\ie each curve in the second column divided by the corresponding curve in the third column)}, which draws attention to the response of the power spectrum that is replicated best and least well by \meadfit. In each case, the cosmological parameter in question is varied over the range that \emu covers, while all other parameters are kept fixed, with the highest value of the cosmological parameter shown in pink and the lowest in blue; the range of each parameter is given in Table~\ref{tab:emu_parameters}. This plot is slightly different from similar plots in \citeauthor{Heitmann2014} (\citeyear{Heitmann2014}) because we use $k/h$ rather than $k$ on the $x$-axis.}
\label{fig:cosmology_variation_0.0}
\end{figure*}

In Fig.~\ref{fig:cosmology_variation_0.0} we show the response of the matter power spectrum to changes in cosmological parameters for the linear spectrum, and the non-linear spectrum predicted by both \meadfit and \emu. This is shown for the cosmological parameters that \emu allows one to vary ($\omega_\mathrm{m}$, $\omega_\mathrm{b}$, $n_\mathrm{s}$, $w$, $\sigma_8$ $h$) over the range that \emu allows (see Table~\ref{tab:emu_parameters}) via the ratio of the power as the cosmological parameter is varied compared to the power for the \newtwo{`fiducial' cosmology, given in Table~\ref{tab:emu_parameters}}. This quantity says nothing about the absolute accuracy of the \meadfit predictions, but allows us to assess if the response of the power induced by changes in cosmological parameters is accurate. Therefore, we also compute the ratio of the response of the \meadfit and \emu predictions (right-hand column). We see that, in general, the matter power spectrum response predicted by \meadfit is in excellent agreement with the simulations of \emu, which is due to the large amount of well-tested theory in the halo model. The most obvious small discrepancies arise around the `bump' at $k\simeq 1\iMpc$ produced by variations in $\sigma_8$ and the degree of enhanced power at $k>1\iMpc$ when $w$ is increased. In each case, the general trend is reproduced well by the halo model but the exact magnitude of the response is not predicted quite correctly.

The accuracy of the predicted halo-model power spectrum depends on the ingredients used. We used Fig. \ref{fig:cosmology_variation_0.0} to inform the ingredients used for \meadfit in order to produce the correct power-spectrum response, before we fitted halo parameters to make accurate $\Delta^2(k)$ predictions. This was particularly important in the case of the response to variations in $w$, which enter our model partly through the concentration--mass relation of \cite{Bullock2001} and partly via the \cite{Dolag2004} correction to a $c(M)$ relation. The vast majority of concentration--mass relations available in the literature produce either no, or too little, response in the matter spectrum to changes in $w$ when the linear theory $\Delta^2(k)$ is held constant.

\section{The halo model calculation}
\label{app:calculation}

In this appendix we discuss some of the practicalities of our approach of calculating the halo-model power spectrum. This appendix serves as a manual to the specific form of the halo-model calculation that our public \textsc{fortran 90} code performs. The halo-model code used in this work is available at \meadaddress.

\meadfit is written such that by default it uses the \cite{Eisenstein1998} approximation to the linear spectrum, which can be used if accuracy at linear scales is not demanded, and then converts this to a non-linear spectrum. If necessary there is the option to read in a tabulated linear spectrum, $k$ versus $P(k)$,  as input (\eg from \textsc{camb}). Additionally, the cosmological parameters need to be specified because they are used in the fitting functions and in calculations of the growth function. In all cases the input linear spectra are renormalized to give the desired $\sigma_8$.

For our calculation of the growth function, we explicitly integrate the approximate expression for the logarithmic growth \emph{rate} given in \cite{Linder2005} and derived in \cite{Linder2007},
\begin{equation}
\frac{\mathrm{d}\ln{g}}{\mathrm{d}\ln{a}}=\Om^\gamma(z)\ .
\end{equation}
where $g$ is the growth factor normalized to be $1$ today, $\gamma=0.55$ if $w=-1$ and $\gamma=0.55+0.02(1+w)$ if $w<-1$ and $\gamma=0.55+0.05(1+w)$ if $w>-1$. This fitting formula and subsequent integration to find the growth factor is valid at the sub-per cent level even if $w$ deviates significantly from $-1$. 

\subsection{Two-halo term}

The two halo term we use is:
\begin{equation}
\Delta^2_\mathrm{2H}(k)=\left[1-f\tanh^2{(k\sigma_\mathrm{v}/\sqrt{f})}\right]\Delta^2_\mathrm{lin}(k)\ ,
\end{equation}
with $f=0.188\times\sigma^{4.29}_8(z)$. The parameter $\sigma_\mathrm{v}$ is calculated via
\begin{equation}
\sigma^2_\mathrm{v}=\frac{1}{3}\int_0^{\infty}\frac{\Delta^2_\mathrm{lin}(k)}{k^3}\;\mathrm{d}\emph{k}\ ,
\end{equation}
and we do this by transforming the $[0,\infty]$ interval in $k$ to $t\in[0,1]$ using the transformation $1+k=1/t$.

The $\sigma_\mathrm{v}$ integral (and that of $\sigma(R)$ in equation~\ref{eq:sig_appendix}) requires knowledge of $\Delta^2$ at very small or very large values of $k$. It is impractical to provide a tabulated linear spectrum over such a large range of scales and so we interpolate beyond the boundaries of the input linear spectrum using the scaling $\Delta^2(k)\propto k^{3+n_\mathrm{s}}$ at small $k$, and the approximate scaling $\Delta^2(k)\propto k^{n_\mathrm{s}-1}\ln^2(k)$ at high $k$. Extremely accurate values of the power at each of these asymptotes are not necessary, because they contribute quite negligibly to the integral, but it is necessary for the linear power not to be badly wrong, or set to zero unphysically.

\subsection{One-halo term}

The aim is to numerically evaluate the integral in equation~(\ref{eq:halopower}). To do this we find it convenient to convert from an integral over $M$ to one over $\nu=\dc/\sigma(M)$:
\begin{equation}
\eqalign{
\Delta_\mathrm{1H}^2(k)=&[1-\epow{-(k/k_*)^2}]4\pi\left(\frac{k}{2\pi}\right)^3\frac{1}{\bar\rho}\cr&\times\int_0^\infty M(\nu) W^2(\nu^\eta k,M) f(\nu)\;\mathrm{d}\nu\ .
}
\label{eq:halopower_nu}
\end{equation}
Here $k_*=0.584\,\sigma_\mathrm{v}^{-1}(z)$ and damps the one-halo term to prevent one-halo power from rising above linear at the largest scales. $\eta=0.603-0.3\,\sigma_8(z)$ and bloats haloes while they maintain a constant virial radius. We integrate over a \emph{finite} range in $\nu$ that captures all haloes necessary to produce a convergent power spectrum at the scales we investigate. In testing we found that $\nu\in[0.3, 5]$ was sufficient to produce convergent results to $k=100\iMpc$ and $\nu\in[0.1, 5]$ if one required power out to $k=10^4\iMpc$. Convergence at higher wave numbers requires the minimum value of $\nu$ to be reduced, because low-mass haloes contribute to the power only at small scales.

The halo-model integral in equation (\ref{eq:halopower_nu}) requires knowledge of $r_\mathrm{v}$, $M$, $c$ and $\sigma$, all as a function of halo mass. In practice, we tabulate these over the finite range in $\nu$ as an `initialization' step to the calculation. We optimized the number of points in $\nu$ in order to produce convergent results up to $k=100\iMpc$. $\sigma(R)$ is computed via
\begin{equation}
\sigma^2(R)=\int_0^{\infty}\Delta^2(k)\, T^2(kR)\;\mathrm{d}\ln{k}\ ,
\label{eq:sig_appendix}
\end{equation}
where
\begin{equation}
T(x)=\frac{3}{x^3}(\sin{x}-x\cos{x})\ .
\end{equation}
We convert the $[0,\infty]$ $k$ range of the integral in equation~(\ref{eq:sig_appendix}) to a finite range in $t$ using $1+k=1/t$ with $t\in[0,1]$. Since this integral is relatively expensive to compute, we generate a look-up table for $\sigma(R)$ with $R$ values logarithmically spaced between $10^{-4}$ and $10^{3}\Mpc$. These radii correspond to all haloes of practical interest at low $z$. If values of $\sigma(R)$ are required outside this range, then we interpolate beyond the table boundaries. This works extremely well because $\sigma(R)$ is a very smooth function in log space with the asymptotes approximating power laws to a very high degree of accuracy for standard cosmological spectra. The radial scale $R$ is related to the mass scale via $M=4\pi R^3 \bar\rho/3$ .

The halo window function in equation~(\ref{eq:halopower_nu}), $W(k,M)$, has an analytical form for an NFW profile \citep{Cooray2002}:
\begin{equation}
\eqalign{
W(k,M)&f(c)=[\Ci(k_\mathrm{s}(1+c))-\Ci(k_\mathrm{s})]\cos(k_\mathrm{s})\cr&
+[\Si(k_\mathrm{s}(1+c))-\Si(k_\mathrm{s})]\sin(k_\mathrm{s})-\frac{\sin(ck_\mathrm{s})}{k_\mathrm{s}(1+c)}\ ,
}
\end{equation}
where $f(c)=\ln(1+c)-c/(1+c)$, $\Si(x)$ and $\Ci(x)$ are the sine and cosine integrals, $k_\mathrm{s}=kr_\mathrm{v}/c$, $c$ is the halo concentration and $r_\mathrm{v}$ is the halo virial radius, related to the halo mass via $M=4\pi r_\mathrm{v}^3\Dv\bar\rho/3$ and $\Dv=418\times\Om^{-0.352}(z)$. The halo concentration is calculated using the prescription of \cite{Bullock2001}, augmented by that of \cite{Dolag2004}:
\begin{equation}
c(M,z)=A\frac{1+z_\mathrm{f}}{1+z}\frac{g_\mathrm{DE}(z\rightarrow\infty)}{g_{\Lambda}(z\rightarrow\infty)}\ , 
\label{eq:concentration}
\end{equation} 
where $A=3.13$. The growth factor correction only applies if $w\neq-1$ and is the asymptotic ratio of growth factors. The formation redshift is then calculated via
\begin{equation}
\frac{g(z_\mathrm{f})}{g(z)}\sigma(fM,z)=\dc\ ,
\label{eq:zf}
\end{equation}
where $f=0.01$. We numerically invert equation~(\ref{eq:zf}) to find $z_\mathrm{f}$ for a fixed $M$. If $z_\mathrm{f}<z$, then we set $c=A$. The mass function we use in equation~(\ref{eq:halopower_nu}) is that of \cite{Sheth1999}:
\begin{equation}
f(\nu)=A\left[1+\frac{1}{(a\nu^{2})^p}\right]\mathrm{e}^{-a\nu^2/2}\ ,
\label{eq:massfunction}
\end{equation}
where $a=0.707$, $p=0.3$, $A=0.2162$ and $\nu=\dc/\sigma(M)$ with $\dc=1.59+0.0314\,\ln\sigma_8(z)$.

\subsection{Full power}

The final expression for the halo-model power spectrum is then
\begin{equation}
\Delta^2(k)=[(\Delta_\mathrm{2H}^{'2})^\alpha+(\Delta_\mathrm{1H}^{'2})^\alpha]^{1/\alpha}\ ,
\end{equation}
where $\alpha=2.93\times 1.77^{n_\mathrm{eff}}$ and
\begin{equation}
3+n_\mathrm{eff}\equiv \left.-\frac{\mathrm{d}\ln\sigma^2(R)}{\mathrm{d}\ln R}\right\vert_{\sigma=1}\ .
\end{equation}
We calculate the derivative numerically from our $\sigma(R)$ look-up table.

\end{document}